\documentclass[twocolumn]{aastex6}
\usepackage{natbib}
\usepackage{color}
\usepackage{hyperref} 

\def    \apjl  		{\rm {ApJL}}

\def    \apjl  		{\rm {ApJL}}

\def	\cm		{\,{\rm {cm}}}
\def	\K		{\,{\rm K}}
\def	\g		{\,{\rm {g}}}
\def	\mum	{\,{\mu \rm{m}}}

\def \bea {\begin{eqnarray}}
\def \ena {\end{eqnarray}}

\def	\ted	{{\tau_{\rm ed}}}
\def	\tH	{{\tau_{\rm H}}}

\def    \bmu    {{\hbox{\boldsym\char'026}}}	

\def    \bomega {{\hbox{\boldsym\char'041}}}	

\def	\bv	{{\bf v}}

\def	\cm	{\,{\rm cm}}

\def	\D	{{\rm D}}

\def	\erg	{\,{\rm erg}}

\def	\g	{\,{\rm g}}
\def	\gas	{\,{\rm gas}}

\def	\H	{{\rm H}}

\def	\Hz	{\,{\rm Hz}}

\def	\s	{\,{\rm s}}

\def	\km 	{\,{\rm km}}
\def	\yr 	{\,{\rm yr}}

\def	\xhat		{\hat{\bf x}}
\def	\yhat		{\hat{\bf y}}
\def	\zhat		{\hat{\bf z}}

\def    \gas     	{{\rm gas}}
\def    \erf     	{{\rm erf}}

\font\mib=cmmib10

\def\bomega{\hbox{\mib\char"21}}

\def\bmu{\hbox{\mib\char"16}}

\begin{document}
\shorttitle{Dust rotational dynamics in shocks}
\shortauthors{Hoang and Tram}
\title{Dust Rotational Dynamics in C-shocks: Rotational Disruption of Nanoparticles by Stochastic Mechanical Torques and Spinning Dust Emission}

\author{Thiem Hoang}
\affil{Korea Astronomy and Space Science Institute, Daejeon 34055, South Korea; \href{mailto:thiemhoang@kasi.re.kr}{thiemhoang@kasi.re.kr}}
\affil{Korea University of Science and Technology, 217 Gajeong-ro, Yuseong-gu, Daejeon, 34113, South Korea}
\author{Le Ngoc Tram}
\affil{LERMA, Observatoire de Paris, \'Ecole Normal Sup\'erieure, PSL Research University, CNRS, Sorbonne Universit\'es,\\ UPMC Univ. Paris 06, F-75231, Paris, France}
\affil{University of Science and Technology of Hanoi, VAST, 18 Hoang Quoc Viet, Hanoi, Vietnam}

\begin{abstract}
Polycyclic aromatic hydrocarbons (PAHs) and nanoparticles are expected to play an important role in many astrophysical processes due to its dominant surface area, including gas heating, chemistry, star formation , and anomalous microwave emission. In dense magnetized molecular clouds where C-shocks are present, PAHs and nanoparticles are widely believed to originate from grain shattering due to grain-grain collisions. The remaining question is whether these nanoparticles can survive in the dense and hot shocked regions, and how to constrain their size and abundance with observations. In this paper, we present a new mechanism to destroy nanoparticles in C-shocks based on centrifugal stress within rapidly spinning nanoparticles spun-up by stochastic atomic bombardment, which is termed rotational disruption. We find that, due to supersonic neutral gas-charged grain drift in C-shocks, nanoparticles can be spun-up to suprathermal rotation by stochastic torques exerted by supersonic neutral flow. The resulting centrifugal stress within suprathermally rotating nanoparticles can exceed the maximum tensile strength of grain material ($S_{\max}$), resulting in rapid disruption of nanoparticles smaller than $a\sim 1$ nm for $S_{\max}\sim 10^{9}\erg\cm^{-3}$. The proposed disruption mechanism is shown to be more efficient than thermal sputtering in controlling the lower cutoff of grain size distribution in C-shocks. We model microwave emission from spinning nanoparticles in C-shocks subject to supersonic neutral drift and rotational disruption. We find that suprathermally rotating nanoparticles can emit strong microwave radiation, and both peak flux and peak frequency increase with increasing the shock velocity. We suggest spinning dust as a new method to constrain nanoparticles and trace shock velocities in dense, shocked regions. 
\end{abstract}
\keywords{ISM: dust, extinction-molecular cloud-shocks}

\section{Introduction\label{sec:intro}}
Polycyclic aromatic hydrocarbons (PAHs) and very small dust grains (hereafter referred to as nanoparticles) are expected to play an important role in gas heating via photoelectric effect \citep{2001ApJS..134..263W} and chemistry because they contribute the dominant surface area of dust grains (see \citealt{2013ApJ...766....8A}). In dense, low ionization molecular clouds, nanoparticles are believed to affect cloud dynamics and star formation process such as ambipolar diffusion (\citealt{1956MNRAS.116..503M}) due to its dominant charge carrier (see \citealt{2016MNRAS.460.2050Z} and references therein). 

Nanoparticles are expected to be depleted in dense regions due to grain coagulation and accretion of atoms on the grain surface (\citealt{2003ARA&A..41..241D}; \citealt{2016MNRAS.456.2290T}). At the same time, grain-grain collisions due to grain acceleration by magnetohydrodynamic (MHD) turbulence (\citealt{2003ApJ...592L..33Y}; \citealt{Hoang:2012cx}) can shatter large grains ($a\gtrsim 0.1\mum$) to form nanoparticles. In a strong radiation field such as massive stars and supernovae, nanoparticles can be produced by rotational disruption of large grains by radiative torques \citep{2018arXiv181005557H}. When a shock is passing through a dense cloud, nanoparticles can also be formed by grain-grain collisions (\citealt{1994ApJ...431..321T}; \citealt{1994ApJ...433..797J}). 

Interstellar shocks are ubiquitous in the interstellar medium (ISM). In magnetized dense clouds with a low ionization fraction, neutral gas and neutral grains can drift relative to charged grains/ions which are coupled to the magnetic field. When the shock speed is lower than the magnetosonic speed, physical parameters are continuous throughout the shock, for which the term C-type shocks are introduced \citep{1980ApJ...241.1021D} (see also \citealt{2004ApJ...610..781C}). As a result, collisions between neutral small grains and charged large grains can shatter the grains and form nanoparticles, including PAHs, nanosilicates, and nanodiamonds (\citealt{1996ApJ...469..740J}; \citealt{2011A&A...527A.123G}). The remaining question is (i) whether nanoparticles can survive passing the C-shock and (ii) how to constrain the size and abundance of this nanodust population in shocked regions.

The first issue is important for understanding the composition of interstellar nanoparticles, which can help to shed light on the exact carrier of anomalous microwave emission (\citealt{2016ApJ...824...18H}; see \citealt{Dickinson:2018ix} for a review). The second issue is important for a quantitative understanding of the effect of nanoparticles on the cloud dynamics and chemistry in interstellar shocks.

Due to the lack of UV/optical photons to trigger mid-IR emission, the potential new technique to probe PAHs and nanoparticles in shocked dense regions is to use microwave emission produced by rapidly spinning nanoparticles (\citealt{1998ApJ...508..157D}; \citealt{Hoang:2010jy}; \citealt{2016ApJ...824...18H}; \citealt{Hoang:2018hc}). Spinning dust emissivity depends mostly on the rotation of nanoparticles, grain dipole moment, and its abundance and size distribution in which the smallest nanoparticles are the most important emitters ( \citealt{2011ApJ...741...87H}). 

Rotational dynamics of interstellar dust grains is a fundamental astrophysical problem. The grain rotation controls a variety of astrophysical observations, both in emission and polarization. For instance, the rapid rotation of nanoparticles can emit electric dipole radiation which is an important galactic foreground contaminant to Cosmic Microwave Background (CMB) radiation. In particular, suprathermal rotation can help dust grains to be efficiently aligned with magnetic fields (\citealt{2016ApJ...831..159H}; see \citealt{Andersson:2015bq} and \citealt{LAH15} for recent reviews). Dust polarization induced by aligned grains is a powerful tracer of cosmic magnetic fields and also a major foreground of CMB polarization. In particular, rotational disruption due to centrifugal stress within extremely fast rotating grains spun-up by radiative torques can control the maximum cutoff of grain size distribution (\citealt{2018arXiv181005557H}). Rotational dynamics of nanoparticles and dust grains has been well studied for the homogeneous diffuse ISM, molecular clouds (\citealt{1998ApJ...508..157D}; \citealt{Hoang:2010jy}), and turbulent media \citep{2011ApJ...741...87H}. Yet a detailed study of grain rotation in interstellar shocks is still not yet available.

The size distribution of PAHs and nanoparticles in the shocked dense regions is poorly known due to the lack of observational constraints. In the diffuse ISM, the smallest size of nanoparticles of $a\sim 0.35$ nm is determined by thermal sublimation (\citealt{1989ApJ...345..230G}; \citealt{2017ApJ...834..134H}). In very hot plasma, thermal sputtering takes over \citep{1979ApJ...231...77D}. In dense cold clouds, due to the lack of UV photons, both sublimation and thermal sputtering are not effective, such that one can expect a much smaller lower cutoff of the grain size distribution. In studies of grain shattering, the smallest size of nanoparticles is usually fixed to $a_{\min}=0.5$ nm without physical justification (\citealt{1996ApJ...469..740J}; \citealt{2011A&A...527A.123G}). \cite{2010A&A...510A..36M} studied the destruction of PAHs in shocks and found that PAHs can be efficiently destroyed by shocks of velocities $v_{s}>100\km\s^{-1}$. For lower shock velocities, PAHs and smallest nanoparticles (i.e., nanoparticles smaller than several nanometers) are expected to survive the shock passage. As a result, constraining the lower size cutoff and abundance of nanoparticles is of great importance. 

In the diffuse ISM, nanoparticles are known to rotate subthermally (i.e., rotational temperature lower than gas temperature) due to the dominance of rotational damping by electric dipole emission over gas collisional damping (\citealt{1998ApJ...508..157D}; \citealt{Hoang:2010jy}; \citealt{2016ApJ...821...91H}). The situation is dramatically different in dense regions where the gas collisional damping becomes dominant over electric dipole emission damping such that nanoparticles can achieve thermal rotation (\citealt{Hoang:2018hc}; see Section \ref{sec:disrupt}). In C-shocks where the gas can be heated to high temperatures of $T_{\gas}\sim 3\times 10^{3}\K$ (see Section \ref{sec:model}), nanoparticles at thermal equilibrium would rotate extremely fast, at frequencies
\bea
\frac{\omega_{\rm rot}}{2\pi}\simeq 2.5\times 10^{10}\left(\frac{T_{\gas}}{3000\K}\right)^{1/2}a_{-7}^{-5/2}\Hz,\label{eq:omega_T}
\ena
where $a$ is the radius of spherical nanoparticles and $a_{-7}=a/(10^{-7}\cm)$ (see Eq. \ref{eq:omega_Trot}). Therefore, nanoparticles of $a\le 0.5$ nm can spin at $\omega_{\rm rot}/2\pi\ge 1.4\times 10^{11}$ Hz, just a factor of 2.6 lower than the critical limit for disruption of $\omega_{\rm cri}/2\pi\sim 3.7\times 10^{11}(0.5~{\rm nm}/a)(S_{\max}/10^{10}\erg\cm^{-3})^{1/2}$ Hz (see Eq. \ref{eq:omega_cri}). As we will see in Sections \ref{sec:rot} and \ref{sec:disrupt}, supersonic neutral drift can further spin-up charged nanoparticles to suprathermal rotation of $\omega>\omega_{\rm cri}$, resulting in an instantaneous disruption of the nanoparticle into tiny fragments. This important effect will be quantified in the present paper.

A high fraction of nanoparticles resulting from grain shattering and suprathermal rotation excited by neutral drift are expected to induce strong spinning dust emission at microwave frequencies. On the other hand, rotational disruption can decrease the abundance of smallest nanoparticles, which can decrease the resulting spinning dust emissivity. We will perform a detailed modeling of spinning dust emission in C-shocks, accounting for this destruction effect.  

The structure of the present paper is as follows. In Section \ref{sec:model}, we describe the C-type shock model in dense magnetized clouds and compute the gas temperature as well as velocities of neutral, ion, and charged nanoparticles. In Section \ref{sec:rot} we will study rotation dynamics of nanoparticles in C-shocks where the supersonic drift of neutrals relative to charged grains is important. We will introduce the rotational disruption mechanism and calculate the minimum size of nanoparticles that can survive the shock passage in Section \ref{sec:disrupt}. In Section \ref{sec:spindust} we calculate spinning dust emissivity from nanoparticles in shocked regions. We will discuss the importance of rotational disruption and potential application of spinning dust for constraining abundance of nanoparticles and shock velocity in Section \ref{sec:discuss}. A short summary of our main results is presented in Section \ref{sec:sum}.

\section{Structures of C-shocks and grain drift velocities}\label{sec:model}
\subsection{Shock structure and physical parameters}
Let us briefly describe the C-type shock in dense magnetized molecular clouds. In the shock reference frame, the ambient pre-shock gas flows into the shock such that their physical parameters change smoothly with the distance in the shock. At the shock interface, the neutral and ion velocities are the same as the shock velocity considered in the shock reference frame. Due to the deceleration when colliding with the shock matter, neutrals and ions are slowed down until they move together with the shock front, i.e., $v_{n}=v_{i}=0$. Due to magnetic forces, ions and charged grains are coupled to the ambient magnetic field and move slower than neutrals, resulting in $v_{n}> v_{i}$ or drift of neutral gas with respect to charged grains and ions. 

We calculate the ion and neutral velocities for the different shock velocities using the one-dimensional plane-parallel Paris-Durham shock model \citep{2015A&A...578A..63F}. Our initial elemental abundances in the gas, grain cores, ice mantle, and PAHs are the same as in the previous studies (\citealt{2003MNRAS.343..390F}; \citealt{2013A&A...550A.106L}; \citealt{2018MNRAS.473.1472T}). 

Different types of shocks depend on the value of the shock's entrance speed relative to the entrance magnetosonic speed $v_{m}$, which is defined in charged fluid as
\bea
v_{m}=\left(c^{2}_{s} + \frac{B^{2}}{4\pi \rho_{c}} \right)^{1/2},\label{eq:vm}
\ena
where $\rho_{c}$ is the total mass density of charged particles, including ions, charged dust grains (e.g., PAHs, nanosilicates, and large grains), $c_{s}$ and $B/\sqrt{4\pi \rho_{c}}$ are the sound speed and Alfv\'en speed of the charged fluid.  The C-shock term is accordingly given when the shock speed is lower than $v_{m}$.

In the case of low ionization (low density of charged particles), Equation (\ref{eq:vm}) yields
\bea
v_{m}\simeq \frac{B}{\sqrt{4\pi \rho_{c}}} = bv_{m1},\label{eq:vm_b}
\ena
where $b$ is a dimensionless magnetization parameter, and $v_{m1}$ is the magnetosonic speed for $b=1$. In our calculations, we find $v_{m1}$=19.2 km$\,$s$^{-1}$ or $v_{m1}$=19.5 km$\,$s$^{-1}$ for gas density of $n_{\H}=10^{4}\,$cm$^{-3}$ or $n_{\H}=10^{5}\,$cm$^{-3}$, respectively. From Equation (\ref{eq:vm_b}) one can obtain the scaling between the magnetic field strength and the gas density:
\bea
B(\mu \rm G)=b \left(\frac{n_{\H}}{1\cm^{-3}}\right)^{1/2},\label{eq:Bfield}
\ena
{where we have assumed homogeneous ionization. For the standard ISM, one has $b\sim 1$, and $b\sim 1-2$ for dense regions driven by outflows (see, e.g., \citealt{2015A&A...575A..98G}). In this paper, we adopt $b=2$, which enables us to explore a wide range of C-shock velocity.}

Table \ref{tab:ISM} presents the main physical parameters, including proton number density $n_{\H}$, gas temperature $T_{\rm gas}$, hydrogen ionization fraction $x_{\H}$, {heavy element ionization fraction $x_{M}$, $\chi$ the radiation strength relative to the average interstellar radiation field (ISRF; \citealt{1983A&A...128..212M}), dust grain temperature $T_{d}$}, molecular hydrogen fraction $y$, fraction of PAHs in the shock $x_{\rm PAH}$. 
\begin{table}
\begin{center}
\caption{Shock Model Parameters}\label{tab:ISM}
\begin{tabular}{l l l l } \hline\hline\\
{\it Parameters} & {Model A} & {Model B} & {Model C}\cr
\hline\\
$v_{\rm s}$(km$\,$s$^{-1}$) &$5-30$ &$5-30$ &$5-30$\cr
$n_{\rm H}(\cm^{-3})$ & $10^{4}$ & $10^{5}$ & $10^{6}$\cr
$T_{\rm gas}$(K)& 10 & 10 & 10 \cr
$T_{\rm d}$(K)& 10 & 10 & 10 \cr
$\chi$ &$0.01$ & 0.01 & 0.01\cr
$x_{\rm H}\equiv n(\H^{+})/n_{\H}$ &$0$ &$0$ &$0$\cr
$x_{\rm M}\equiv n(\rm M^{+})/n_{\H}$ &$10^{-6}$ &$10^{-6}$ &$10^{-6}$\cr
$x_{\rm PAH}\equiv n(\rm PAH)/n_{\H}$ &$10^{-6}$ &$10^{-6}$ &$10^{-6}$\cr
$y=2n(\H_{2})/n_{\H}$ & {$0.999$} &{$0.999$}&{$0.999$}\cr
$B(\mu G)=bn_{\H}^{1/2}$ $^a$ & 200 & 632 & 2000\cr
\hline
\cr
\multicolumn{4}{l}{$^{a}$~{\it Notes}: Here b=2 is assumed.}\cr
\hline\hline
\end{tabular}
\end{center}
\end{table}

\begin{figure}
\includegraphics[width=0.45\textwidth]{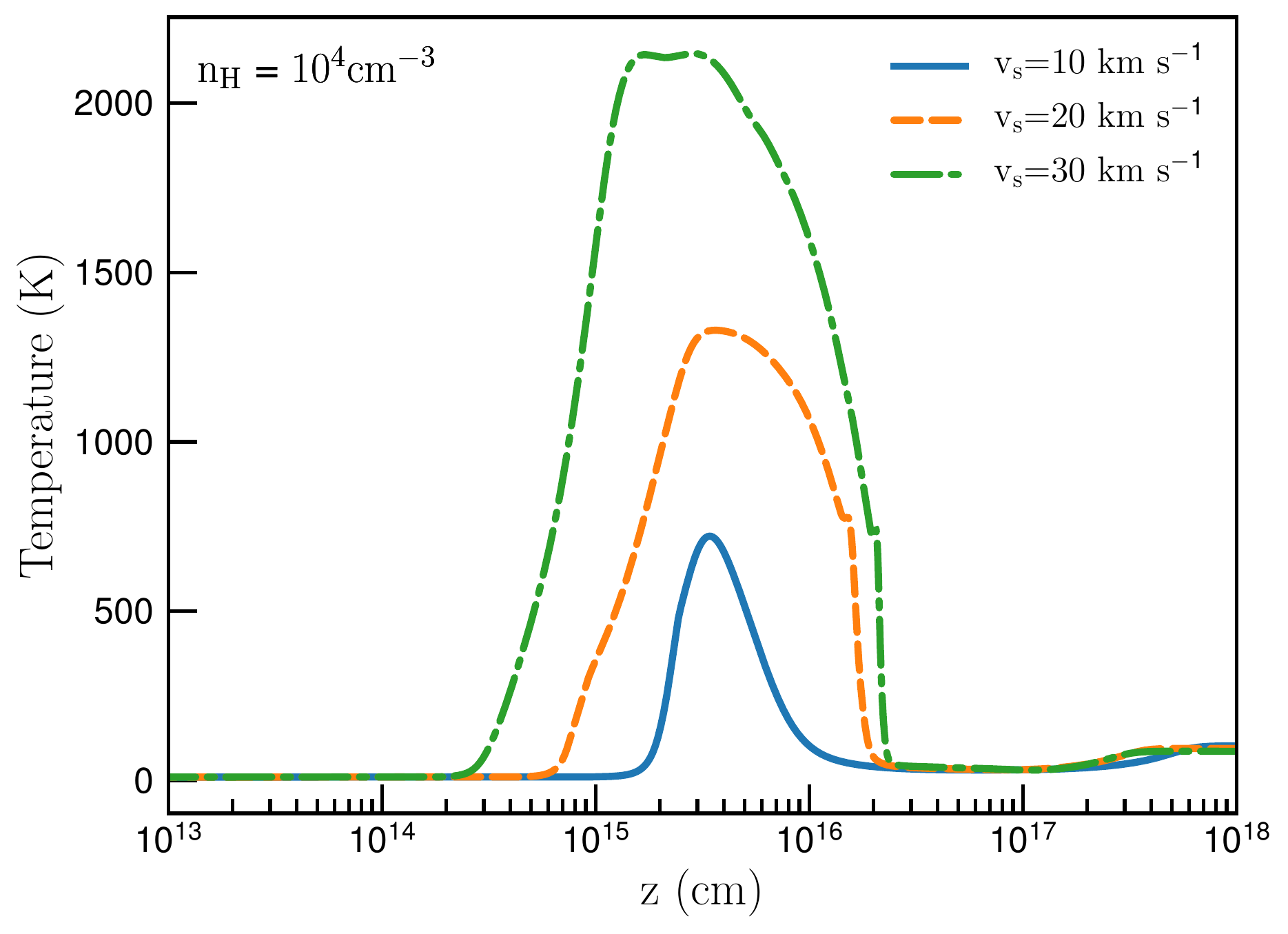}
\includegraphics[width=0.45\textwidth]{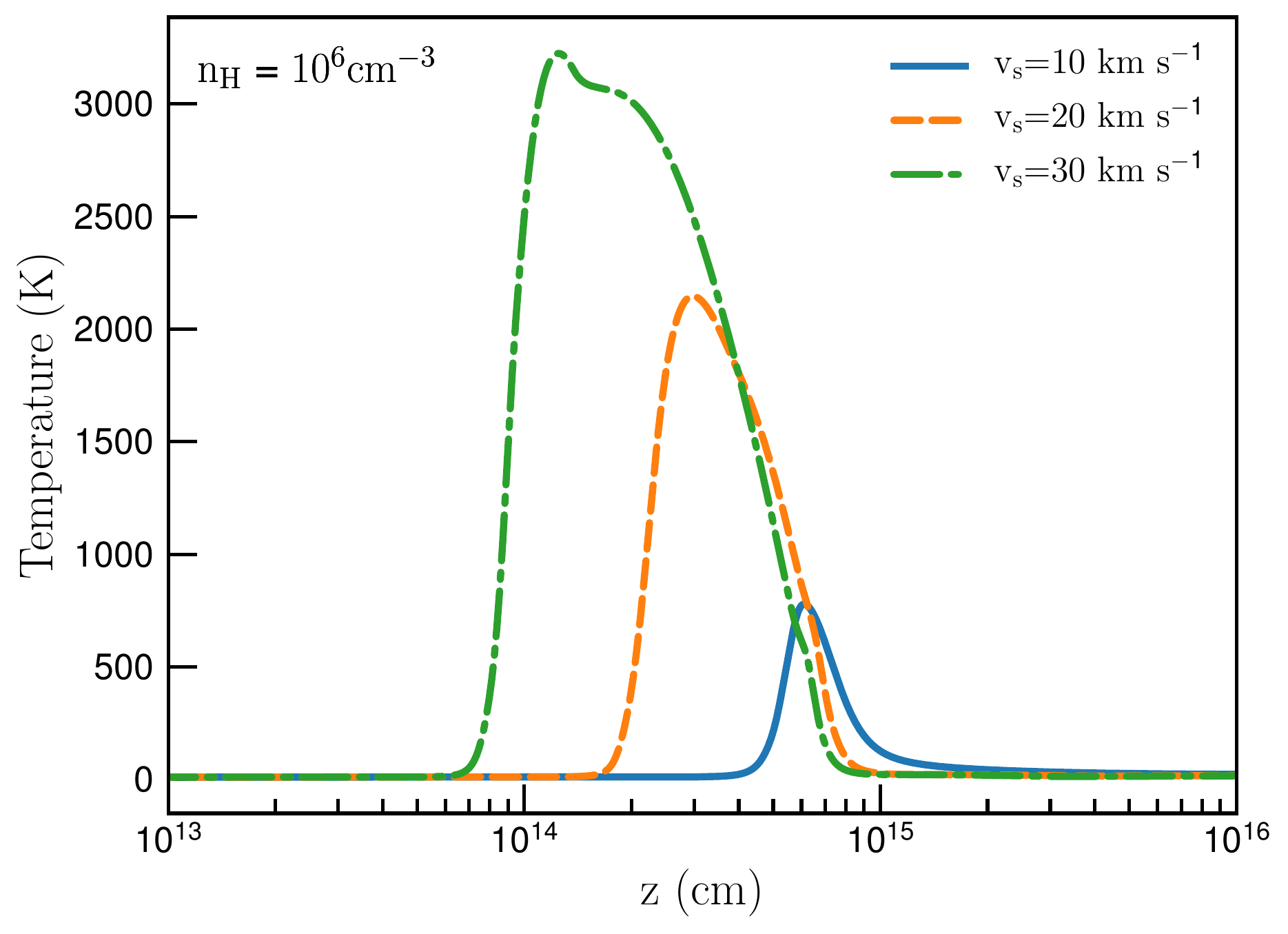}
\caption{Gas temperature vs. distance $z$ in the C-shock of three different shock velocities, assuming $n_{\H}=10^{4}$ cm$^{-3}$ (upper panel) and $n_{\H}=10^{6}$ cm$^{-3}$ (lower panel). The gas temperatures increases with $v_{s}$ and $n_{\H}$.}
\label{fig:Cshock-temp}
\end{figure}

Figure \ref{fig:Cshock-temp} shows the temperature structure of the C-shock for gas density $n_{\H}=10^{4}\cm^{-3}$ (upper panel) and $n_{\H}=10^{6}\cm^{-3}$ (lower panel). Here $z=0$ corresponds to the location at the interface of the preshock and postshock medium. A positive value of $z$ corresponds to the location inside the shock, which is also the shocked gas. Shock compresses the gas and makes the gas temperature to first rise to the peak and then it decreases due to radiative cooling and infrared emission. {The magnetic precursor forms upstream of the discontinuity}, where the charged and neutral fluids dynamically decouple (Figure \ref{fig:Cshock-temp}). The resulting friction between the two fluids heats up and accelerates the neutral fluid. Because of friction between the neutral and charged components, the kinetic energy dissipation is a much more gradual process and spreads over a much larger volume. For instance, the length $L$ of the C-type shock can be up to $3\times 10^{16}\cm$, or $\sim$ 0.01 pc with pre-shock density $n_{\H}=10^{4}\,$cm$^{-3}$. In a denser cloud (lower panel), however, the gas is swept and compressed by shock much stronger and earlier. {This therefore makes the gas hotter, and the gas temperature increases much earlier compared to a less dense cloud}. Radiative cooling also occurs faster, narrowing the shocked region to $L\sim 10^{15}\cm$ (see lower panel).

\subsection{Drift velocities}

\begin{figure}
\includegraphics[width=0.45\textwidth]{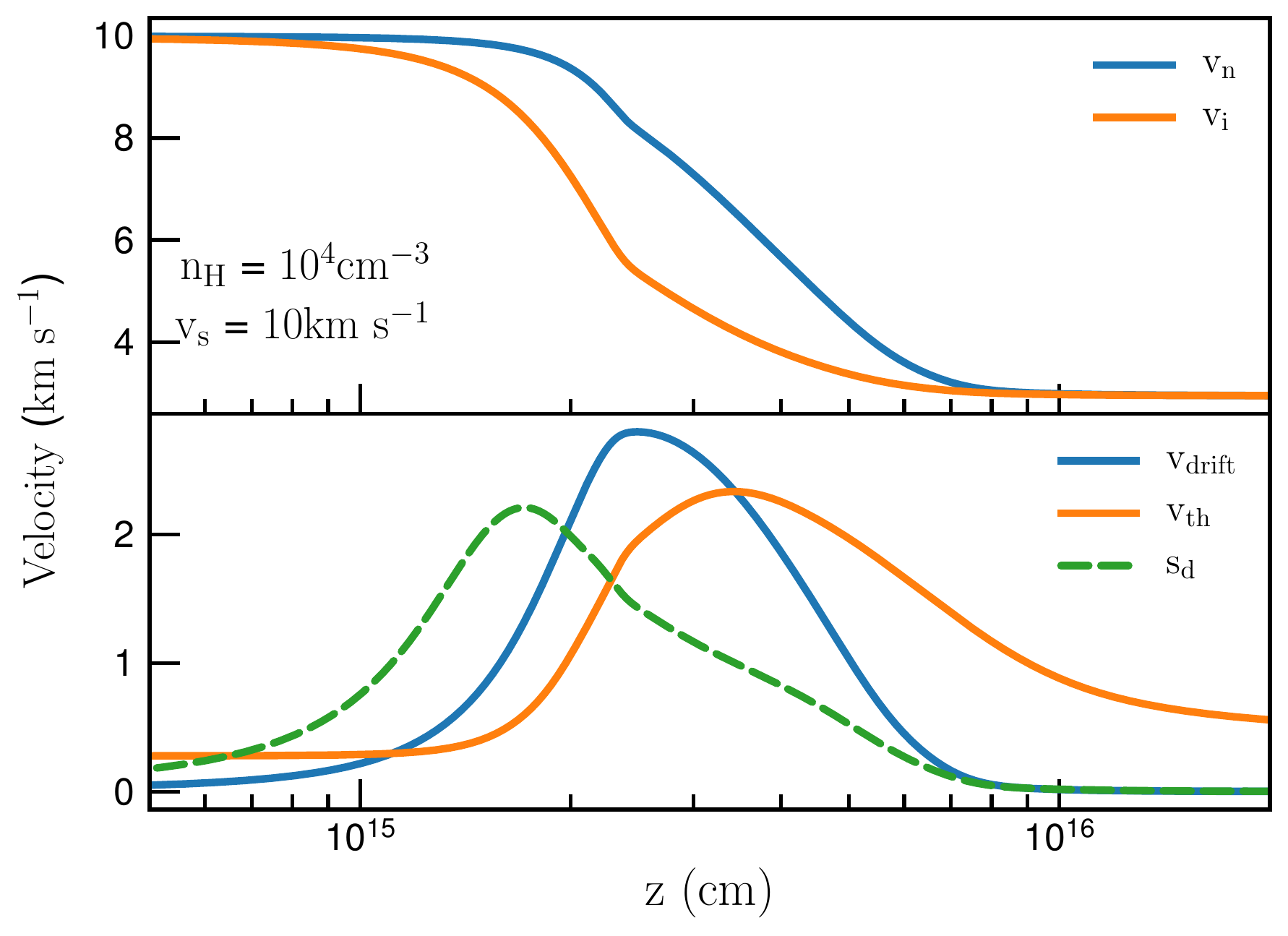}
\includegraphics[width=0.45\textwidth]{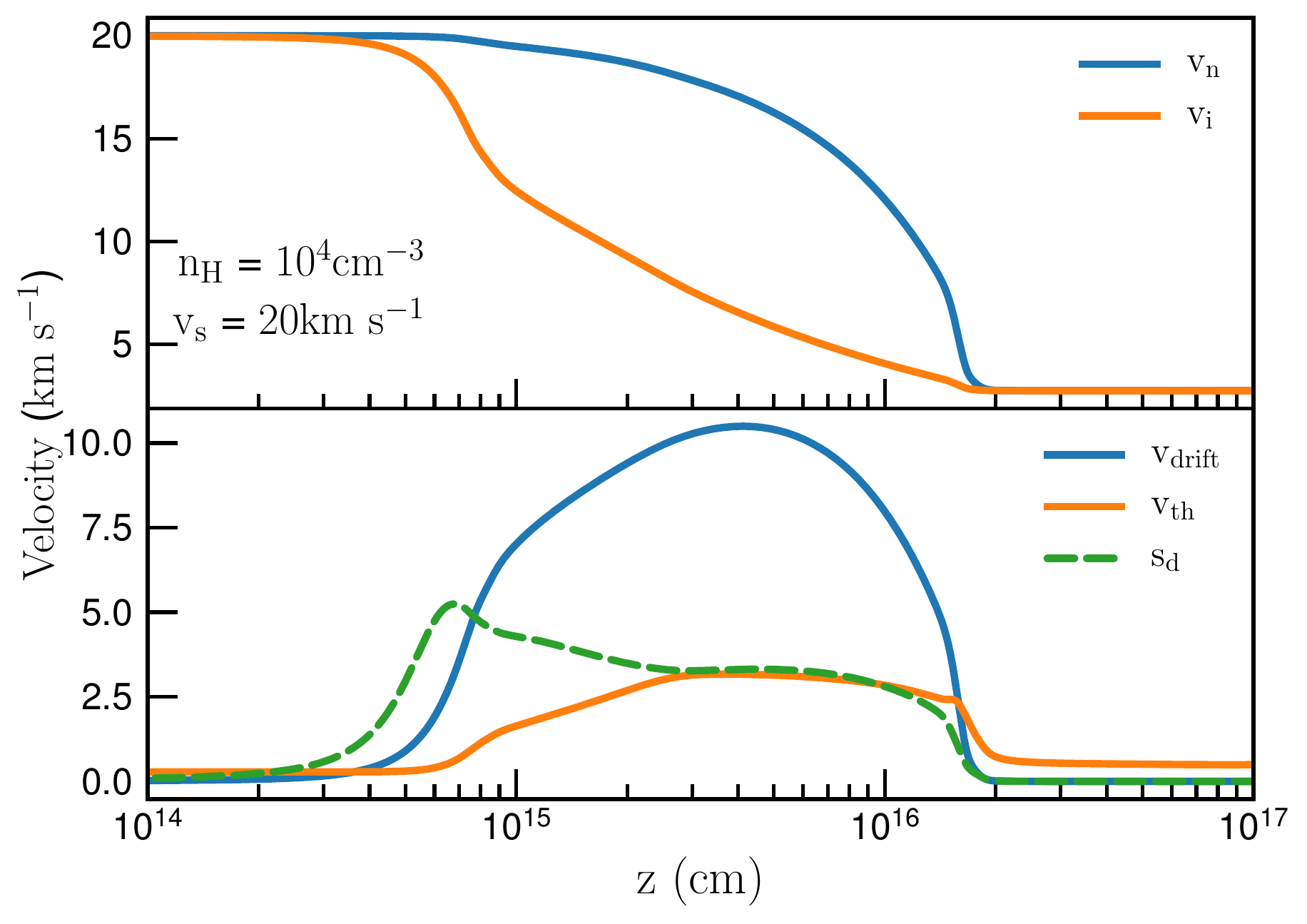}
\includegraphics[width=0.45\textwidth]{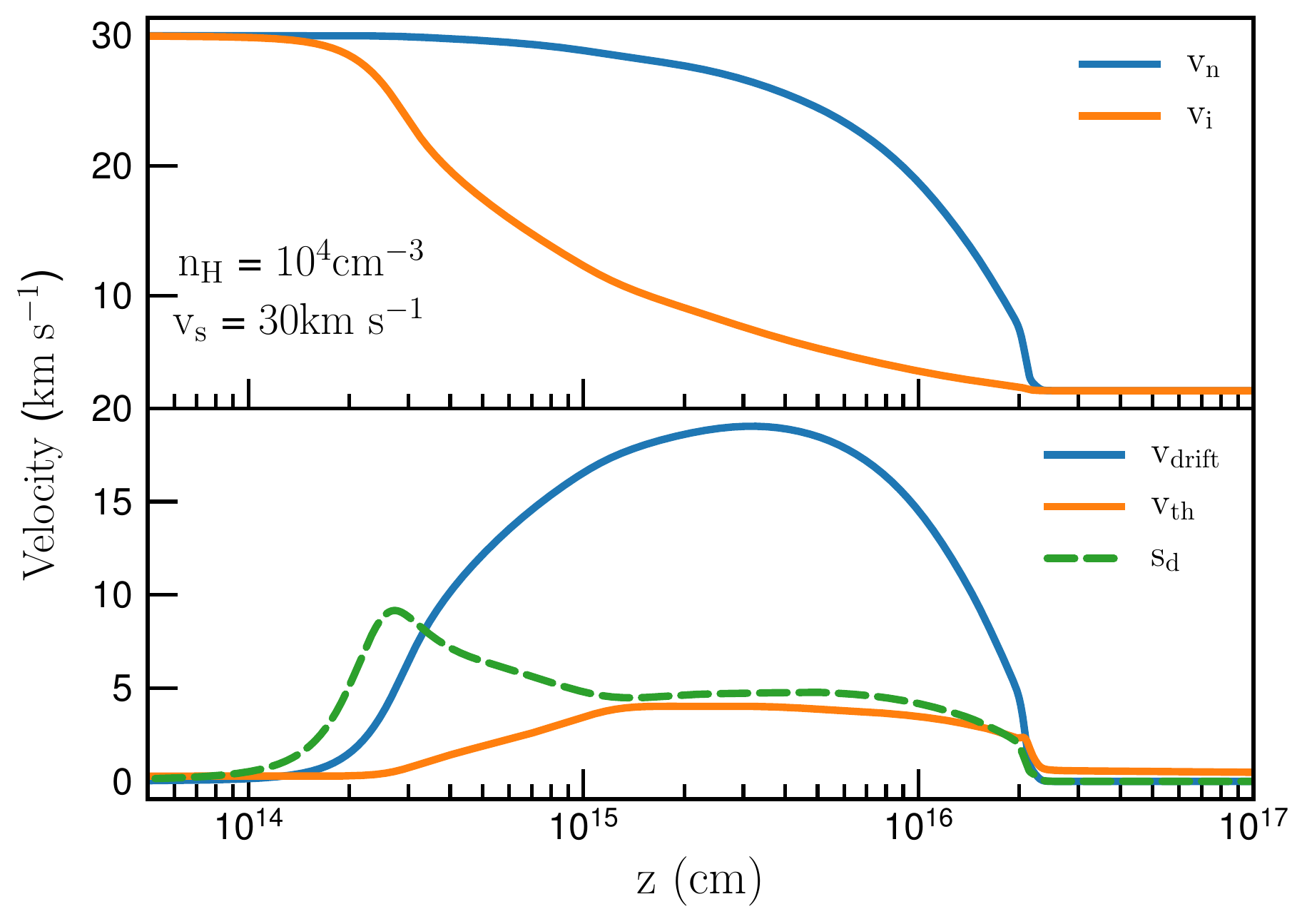}
\caption{Velocity profiles of neutral ($v_{n})$, ion and charged grains ($v_{i}$), and their relative velocity ($v_{\rm drift}=v_{n}-v_{i})$ in the C-shocks with the same parameters as Figure \ref{fig:Cshock-temp}. {The dashed line shows the drift parameter $s_{d}=v_{\rm drift}/v_{th}$ which is dimensionless.} The drift velocity increases with the shock velocity, but $s_{d}$ is slightly changed due to an increased thermal velocity.}
\label{fig:Cshockn4-velo}
\end{figure}

Figures \ref{fig:Cshockn4-velo} shows the velocity structure of neutral ($v_{n}$) and ions ($v_{i}$), {as well as the drift velocity of neutrals relative to ions ($v_{\rm drift}=v_{n}-v_{i}$)}, assuming $n_{\H}=10^{4}\cm^{-3}$. The dimensionless drift parameter is defined as $s_{d}=v_{\rm drift}/v_{\rm th}$ where $v_{\rm th}=\left(2kT_{\rm gas}/m_{\rm H}\right)^{1/2}$ is the thermal gas velocity. 

The drift velocity $v_{\rm drift}$ rises and reaches the maximum value at the middle of the shock and then declines to zero. The drift parameter, $s_{d}$, increases rapidly with $z$, and then declines when the gas is heated to high temperatures. Note the peak of $s_{d}$ does not coincide with the peak of $v_{\rm drift}$ due to the effect of $v_{\rm th}$ or $T_{\rm gas}$. 

\begin{figure}
\includegraphics[width=0.45\textwidth]{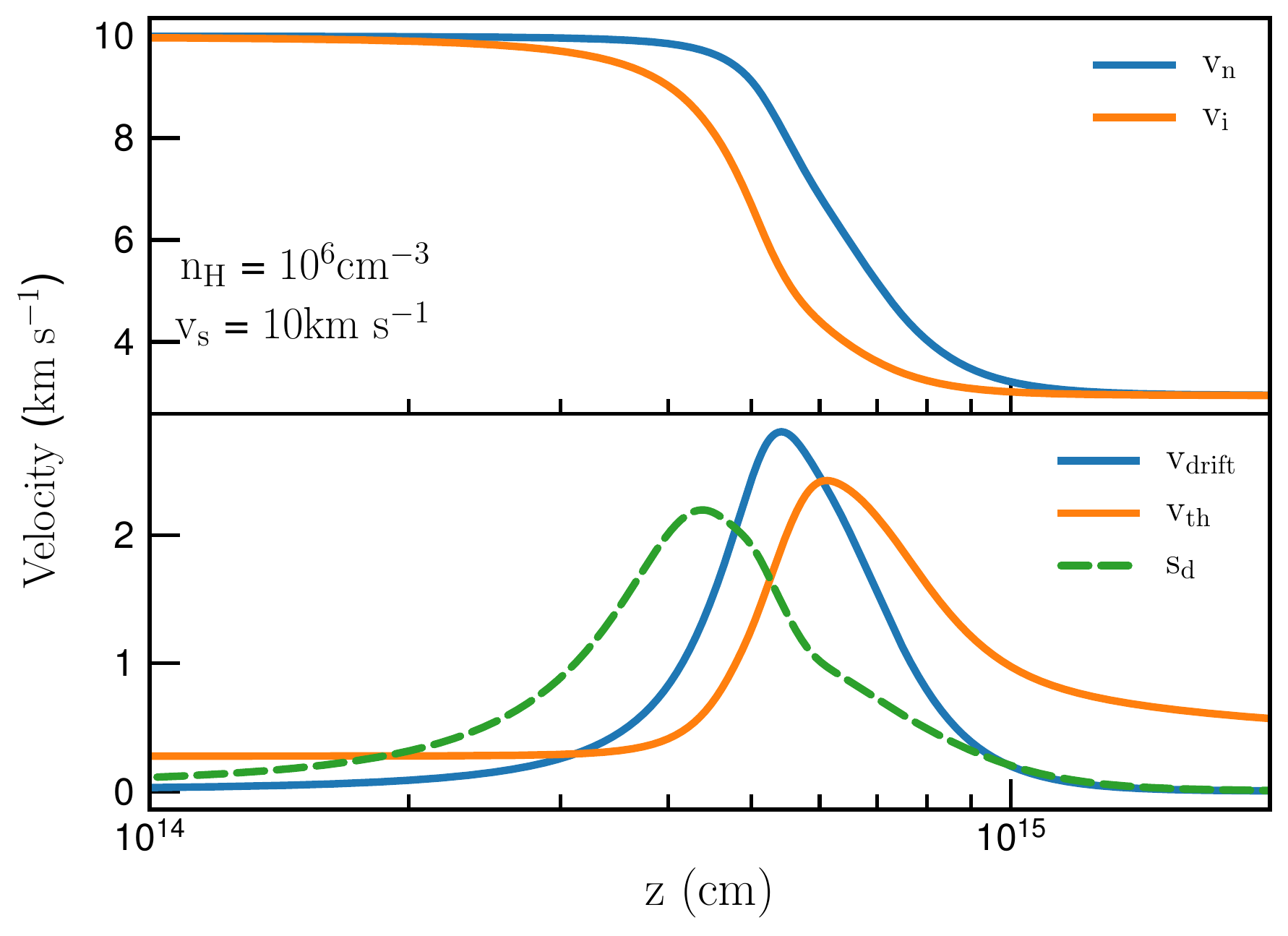}
\includegraphics[width=0.45\textwidth]{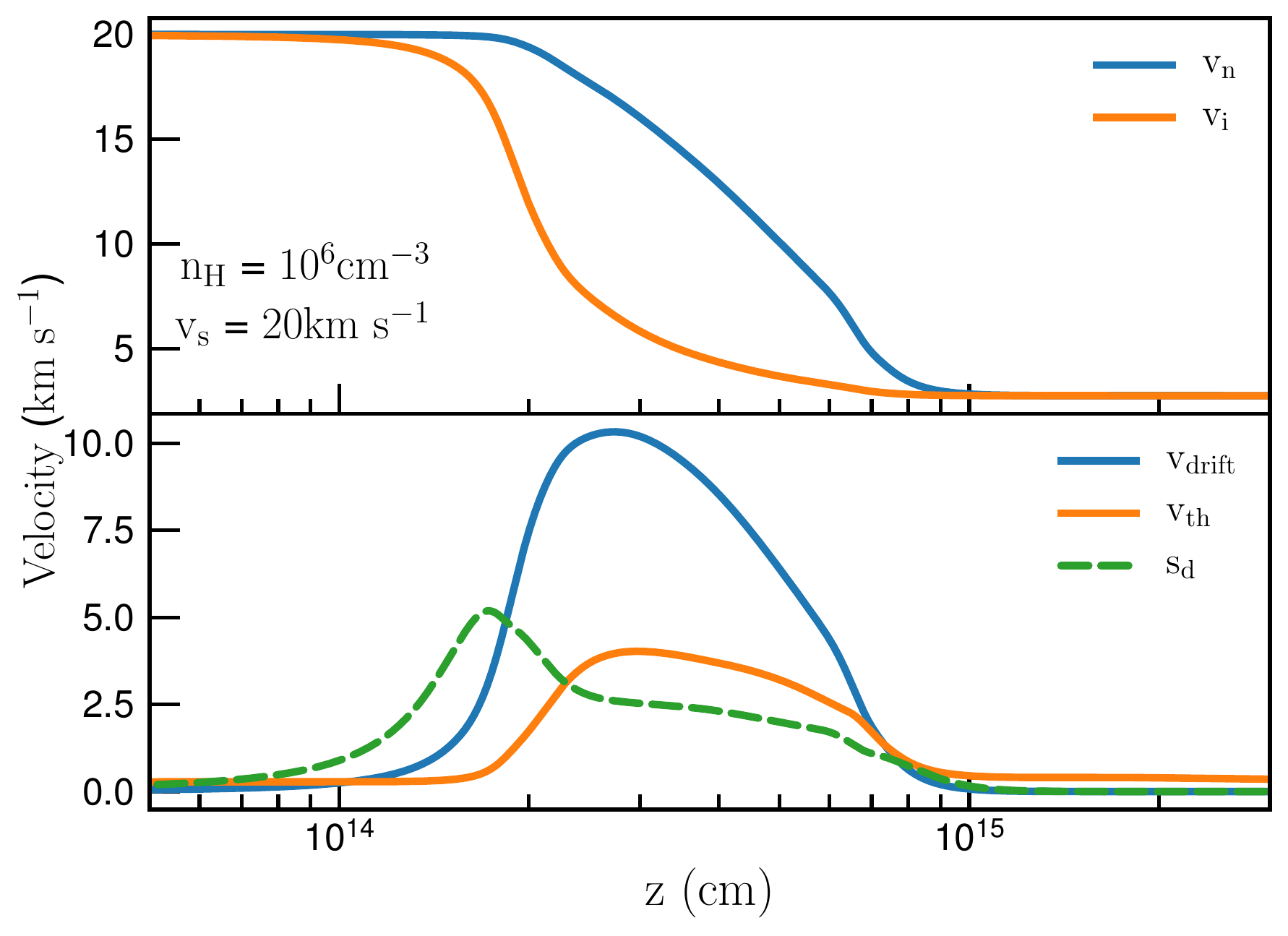}
\includegraphics[width=0.45\textwidth]{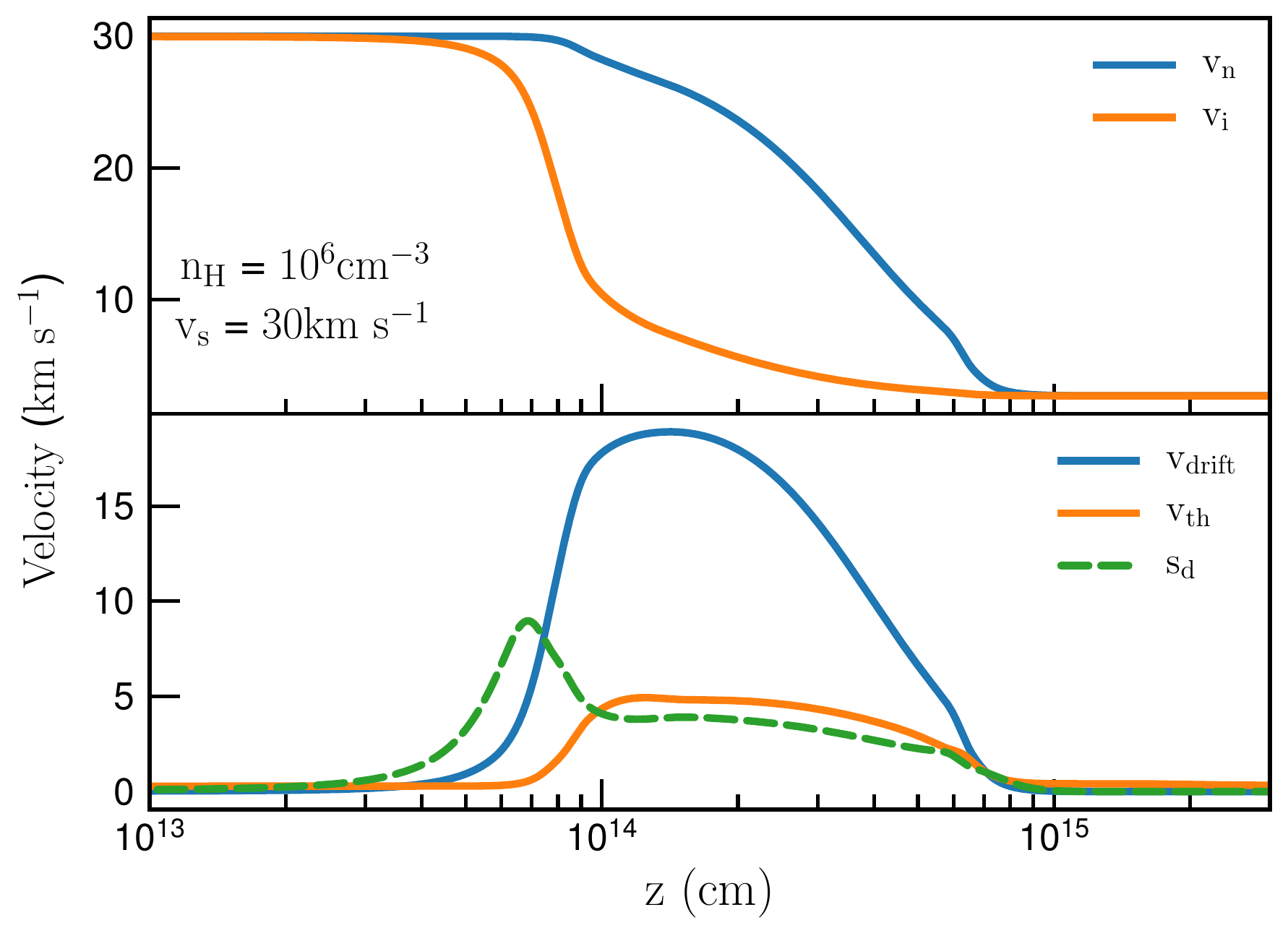}
\caption{Same as Figure \ref{fig:Cshockn4-velo} but with $n_{\H}=10^{6}\cm^{-3}$. Higher thermal and drift velocities but slightly lower values of $s_{d}$ ({dimensionless drift parameter}) are observed.}
\label{fig:Cshockn6-velo}
\end{figure}

Figure \ref{fig:Cshockn6-velo} shows the results for $n_{\H}=10^{6}\cm^{-3}$. Compared to Figure \ref{fig:Cshockn4-velo} ($n_{\H}=10^{4}\cm^{-3}$), gas temperatures are much higher as a result of collisional heating. The shock length is also narrower due to faster radiative cooling.

\section{Rotational dynamics of dust grains in C-type shocks}\label{sec:rot}
\subsection{Collisions with purely atomic gas}
In dense, low ionization regions, collisions with gas atoms and molecules play an important role in grain rotational dynamics. We first consider rotational damping and excitation induced by collisions of purely hydrogen thermal gas with a spherical grain. This situation is the same as those derived in \cite{Jones:1967p2924}, but here we describe in more details for convenience. The treatment below is also valid for molecular gas with $n_{\H}=2n(\H_{2})$.

The well-known damping process for a rotating grain is sticking collision with gas atoms, followed by thermal evaporation. Let consider a grain rotating along the z-axis with angular velocity $\omega_{z}$. The angular momentum carried away by an H atom from the grain surface is given by
\bea
\delta J_{z} = I_{m}\omega_{z} = m_{\H}r^{2}\omega_{z}= m_{\H}a^{2}\sin^{2}\theta\omega_{z},
\ena
where $r$ is the distance from the atom to the spinning axis $z$, $I_{m}=m_{\H}r^{2}$ is the inertial moment of the hydrogen atom of mass $m_{\H}$, $\theta$ is the angle between the z-axis and the radius vector, and $r=a\sin\theta$ is the projected distance to the center. In an interval of time, there are many atoms leaving the grain surface from the different location, $\theta$. Then, assuming the isotropic distribution of $\theta$ the atoms leaving the grain, we can evaluate the mean angular momentum carried away per H atom. Thus, we can replace $\sin^{2}\theta = <\sin^{2}\theta> = 2/3$, which give rise to

\bea
\langle \delta J_{z}\rangle = \frac{2}{3}m_{\H}a^{2}\omega_{z}.
\ena

Using the collision rate of atomic gas, $R_{\rm coll} = n_{\H}v \pi a^{2}$, one can derive the mean decrease of grain angular momentum per unit of time is
\bea
\langle \Delta J'_{z}\rangle &\equiv& \bigg\langle\frac{\Delta J_{z}}{\Delta t}\bigg\rangle_{\H} \nonumber\\
&=& -R_{\rm coll}\langle \delta J_{z}\rangle =-\frac{2}{3}n_{\H}m_{\H}\pi a^{4}\omega_{z}\langle v\rangle.
\ena

The mean velocity is defined by
\bea
\langle v \rangle = Z \int v 4\pi v^{2}e^{-m_{\H}v^{2}/2kT_{\gas}} dv = \left(\frac{8kT_{\gas}}{\pi m_{\H}} \right)^{1/2}, 
\ena
where $Z=(m_{\H}/2\pi kT_{\gas})^{3/2}$ is the normalization factor of the Boltzmann distribution of gas velocity.

Thus, the gas damping rate is
\bea
\langle \Delta J'_{z}\rangle &=&-\frac{2}{3}n_{\H}m_{\H}\pi a^{4}\omega_{z}\left(\frac{8kT_{\gas}}{\pi m_{\H}} \right)^{1/2}\\
&=&-\frac{I\omega_{z}}{\tau_{\H}},
\ena
where $I=8\pi \rho a^{5}/15$ is the inertia moment of a spherical grain with $\rho$ being the dust mass density, and the characteristic rotational damping time $\tau_{\H}$ reads
\bea
\tau_{\H}&=&\frac{3}{4\sqrt{\pi}}\frac{I}{n_{\rm H}m_{\rm H}
v_{\rm th}a^{4}}\nonumber\\
&\simeq& 1.05\times 10^{3}a_{-7}\hat{\rho}\left(\frac{30\cm^{-3}}{n_{\H}}\right)\left(\frac{100\K}{T_{\gas}}\right)^{1/2}~{\rm yr}.\label{eq:taugas}
\ena

Now, let us estimate the rotational excitation of grains due to atomic bombardment.\footnote{We note that this process was first proposed by \cite{1952MNRAS.112..215G} as a mechanism to explain the alignment of dust grains with the magnetic field. Here we extend it for nanoparticles in shocked regions.} Each atom colliding with the grain surface at radius ${\bf r}$ induces an {\it impulsive torque} of ${\bf r}\times m_{\H}\bv$, such that the increase of $(\delta J)^{2}$ from each impact becomes
\bea
(\delta J)^{2} = (rm_{\H}v)^{2}= m_{\H}^{2}v^{2}r^{2}.\label{eq:dJsqr}
\ena
When averaging the the grain surface, one has $\langle r^{2}\rangle= a^{2}/2$, and Equation (\ref{eq:dJsqr}) becomes
\bea
(\delta J)^{2} = \frac{1}{2}m_{\H}^{2}v^{2}a^{2}.
\ena

Using the random walk theory for stochastic collisions, one can derive the total increase of $(\delta J)^{2}$ per unit of time as follows:
\bea
\frac{(\Delta J)_{in}^{2}}{\Delta t} = R_{\rm coll}(\delta J)^{2} = n_{\H}v \pi a^{2}m_{\H}^{2}v^{2}a^{2}/2.\label{eq:dJ2}
\ena

Assuming that thermal evaporation of atoms contributes the same amount of rotational energy as impinging atoms, then, one obtains 
\bea
\frac{(\Delta J)^{2}}{\Delta t}=\frac{2(\Delta J)_{in}^{2}}{\Delta t}=\pi a^{2}n_{\H}m_{\H}^{2}v^{3}a^{2}, \label{eq:DeltaJ2}
\ena
which describes the increase of grain rotational kinetic energy per second.

For thermal gas, the mean rate of energy increase is obtained by integrating Equation (\ref{eq:DeltaJ2}) over the velocity distribution of gas atoms, which results in
\bea
\langle (\Delta J')^{2}\rangle_{\H} \equiv \bigg \langle\frac{(\Delta J)^{2}}{\Delta t}\bigg\rangle_{\H}=\pi a^{2}n_{\H}m_{\H}^{2}\langle v^{3}\rangle a^{2},\label{eq:DeltaJ2_v}
\ena
where 
\bea
\langle v^{3}\rangle &=& C\int v^{3} 4\pi v^{2} e^{-m_{\H}v^{2}/2kT_{\gas}} dv\nonumber\\
&=&\frac{4}{\pi^{1/2}}\left(\frac{2kT_{\gas}}{m_{\H}}\right)^{3/2},\label{eq:vthird}
\ena
with $C=(2\pi kT_{\gas}/m_{\H})^{-3/2}$ being the normalization constant.

Plugging Equations (\ref{eq:vthird}) into (\ref{eq:DeltaJ2_v}) one obtains the {\it excitation coefficient}:
\bea
\langle (\Delta J')^{2}\rangle_{\H}
=n_{\H}(8kT_{\gas}/\pi m_{\H})^{1/2} 4\pi a^{4}m_{\H}kT_{\gas}.
\ena

The one-dimensional excitation coefficient 
\bea
\langle (\Delta J'_{x})^{2}\rangle_{\H}&=&\langle (\Delta J)^{2}\rangle/3=\frac{8kT_{\gas}}{3}(2\pi m_{\H}kT_{\gas})^{1/2} a^{4}n_{\H}\nonumber\\
&=&\frac{(I\omega_{T})^{2}}{\tau_{\H}},\label{eq:deltaJx2}
\ena
where $\omega_{T}=(2kT_{\gas}/I)^{1/2}$ is the thermal angular velocity.

\subsection{Ion collisions, plasma drag, and infrared emission}
Realistic gas consists of atomic and molecular hydrogen as well as heavier elements. In addition to neutral-grain collisions, nanoparticles are directly bombarded by ions and experience long-distant interaction with passing ions (i.e., plasma drag) (\citealt{1998ApJ...508..157D}; \citealt{Hoang:2010jy}). 

Following \cite{1998ApJ...508..157D}, one can define the dimensionless damping ($F$) and excitation ($G$) coefficients for an interaction process with respect to the damping and excitation coefficients of purely hydrogen neutral-grain collisions. The total damping and excitation rate from various interaction processes are given by
\bea
\frac{d(I\omega_{z})}{dt} &=& \langle \Delta J'_{z}\rangle _{\H}\times \sum_{j}F_{j},\\
\frac{d (I\omega^{2})}{dt} &=& \frac{\langle (\Delta J')^{2}\rangle _{\H}}{I}\times G=\frac{3I\omega_{T}^{2}}{\tau_{\H}}\times \sum_{j}G_{j},\label{eq:Gcoeff}
\ena
where $j=n,ion,p,IR$ denotes neutral-grain, ion-grain, plasma drag, and IR emission. Note $F_{j}=G_{j}=1$ for grain collisions with purely atomic hydrogen gas.

The diffusion coefficients from ion-grain collisions and plasma drag can be calculated assuming stationary grains, i.e., $s_{d}=0$, because ions and charged grains are both coupled to the magnetic field.

\subsection{Rotational damping and excitation in C-shocks}
We move on to study rotational dynamics of nanoparticles in C-shocks in the presence of supersonic neutral drift relative to charged grains and derive rotational damping and excitation coefficients.

\subsubsection{Supersonic drift}
For the supersonic drift of $s_{d}\gg 1$, one can easily obtain the total excitation coefficient as given by Equation (\ref{eq:DeltaJ2}): 
\bea
\langle (\Delta J')^{2}\rangle=n_{\H}m_{\H}^{2}v_{\rm drift}^{3}\pi a^{4}.\label{eq:deltaJ2_sup}
\ena
From Equation (\ref{eq:Gcoeff}) and (\ref{eq:deltaJ2_sup}), one obtains the dimensionless excitation coefficient:
\bea
G_{n}(s_{d})=\frac{\pi^{1/2}s_{d}^{3}}{4}.\label{eq:Gsd_sup}
\ena

\subsubsection{Transonic drift}
For the transonic case (i.e., $s_{d}\sim 1$), to compute the diffusion coefficients, we follow the similar approach as in \cite{1995ApJ...453..238R}. Let $\xhat\yhat\zhat$ be the reference frame fixed to the gas, such that $\zhat$-axis is directed along the magnetic field and the drift velocity $v_{d}$ lies in $\yhat\zhat$ plane with an angle $\alpha$ with $\zhat$. In this paper, we consider a perpendicular shock such that $\alpha=90^{\circ}$. For the interest of grain rotation, we here consider only spherical grains. Interaction of a supersonic gas flow with non-spherical grains are important for grain alignment (\citealt{2018ApJ...852..129H}) will be addressed in a future study.

Following \cite{1995ApJ...453..238R}, the dimensionless damping coefficient is given by
\bea
\langle \Delta j_{i}\rangle=-M_{0}j_{i} {~\rm for~} i=x,y,z,
\ena
where $M_{0}$ is a function of drift velocity, and $M_{0}=1$ for $s_{d}=0$, i.e., thermal collision and evaporation.

The diffusion coefficients parallel (P) and transverse (T) to the neutral gas flow are give by ({see Eq. 3.14-3.15 in \citealt{1995ApJ...453..238R}}):
\bea
D_{T}(s_{d}) &=& \frac{3(1 + 2s_{d}^{2})}{4}M_{0}(s_{d}) + (1 - 2s_{d}^{2})M_{2}(s_{d}),\\
D_{P}(s_{d}) &=& \frac{3}{2}\left[M_{0}(s_{d}) - M_{2}(s_{d})\right],
\ena
where
\bea
M_{0} &=& \left(\frac{\sqrt{\pi}}{4s_{d}}\right)\left[2(1 + s_{d}^{2})\erf(s_{d}) - P(3/2,s_{d}^{2})\right], \\
M_{2} &=&  \left(\frac{\sqrt{\pi}}{4}\right)s_{d}\erf(s_{d}) -(\frac{3\sqrt{\pi}}{16})s_{d}^{-3}P(5/2,s_{d}^{2}) \nonumber\\
&&+ \frac{\sqrt{\pi}}{4}s_{d}^{-3}P(3/2,s_{d}^{2})
\ena
{which are given by Equations (3.9) and (3.16) in \cite{1995ApJ...453..238R}, and $P$ is the incomplete gamma function.}

In the reference frame fixed to the ambient gas, the dimensionless diffusion coefficients become
\bea
\langle (\Delta j_{x})^{2}\rangle&=&D_{T}+\frac{T_{d}}{T_{\rm gas}}M_{0}(s_{d}),\label{eq:deltajx}\\
\langle (\Delta j_{y})^{2}\rangle&=&\sin^{2}\alpha D_{P}+\cos^{2}\alpha D_{T}+ \frac{T_{d}}{T_{\rm gas}}M_{0}(s_{d}),\\ 
\langle (\Delta j_{z})^{2}\rangle&=&\cos^{2}\alpha D_{P}+\sin^{2}\alpha D_{T}+ \frac{T_{d}}{T_{\rm gas}}M_{0}(s_{d}),\label{eq:Gsd}
\ena
which results in the dimensionless excitation coefficient
\bea
G_{n}(Z\ne 0)=\frac{1}{3}\sum_{i=x,y,z}\langle (\Delta j_{i})^{2}\rangle.\label{eq:Gn_sd}
\ena

For $s_{d}\gg 1$, $D_{T}$ scales as $s_{d}^{3}$, such that $G_{n}$ returns to Equation (\ref{eq:Gsd_sup}).

\subsection{Dynamical timescales in shocks}\label{sec:times}
For grain dynamics in shocks, an important dynamical timescale is the flow time of dust grains in the C-shock. The grain flow time at location $z$ in the shock can be found by
\bea
\tau_{\rm flow}=\int_{0}^{z}\left(\frac{dz' }{v_{\rm drift}(z')}\right),\label{eq:tflow}
\ena
where the drift velocity is a function of distance. Assuming $v_{\rm drift}\sim 10 \km\s^{-1}$ for the shock length $L=10^{15}\cm$, we can estimate the passage time as
\bea
\tau_{\rm flow}=\frac{L}{v_{\rm drift}}=30\left(\frac{L}{10^{15}\cm}\right)\left(\frac{10\km \s^{-1}}{v_{\rm drift}}\right)\rm yr.\label{eq:tflow}
\ena

To understand the effect of rotational excitation by gas bombardment, we need to compare the spin-up time by stochastic collisions with the flow time. The characteristic timescale to spin up a grain from the rest to an angular momentum $J$:
\bea
\tau_{\rm spin-up}=\frac{J^{2}}{(\Delta J)^{2}/(\Delta t)}
=\frac{J^{2}}{n_{\H}m_{\H}^{2}v_{\rm drift}^{3}\pi a^{4}},\label{eq:tspin}
\ena
where Equation (\ref{eq:DeltaJ2}) has been used and $v_{\rm drift}$ is the velocity of the neutral drift.

Assuming $J=I\omega_{\rm T}$, the spin-up time by neutral gas drift is equal to
\bea
\tau_{\rm spin-up}&=&\frac{16\rho k T_{\gas} a}{15n_{\H} m_{\H} ^{2}v_{\rm drift}^{3}} \nonumber\\
&=&0.005a_{-7}\left(\frac{T_{\gas}}{10^{3}\K}\right)
\left(\frac{n_{\H}}{10^{5}\cm^{-3}}\right)^{-1}\nonumber\\
&&\times\left(\frac{v_{\rm drift}}{10\km\s^{-1}}\right)^{-3}\yr.\label{eq:tspinup}
\ena

By comparing Equations (\ref{eq:tflow}) with (\ref{eq:tspinup}), it follows that the spin-up time by Gold stochastic torques is rather short compared to the time passing the shock structure. Therefore, nanoparticles can be rotating suprathermally in the substantial fraction of the shock. The spin-up timescale is larger for larger grains, such as classical grains of $a=0.1\mum$ because $\tau_{\rm spin-up}\propto a$.

The rotational damping time is given by Equation (\ref{eq:taugas}), which corresponds to 
\bea
\tau_{\rm H}\simeq 0.067 a_{-7}\left(\frac{T_{\gas}}{1000\K}\right)^{-1/2}
\left(\frac{n_{\H}}{10^{5}\cm^{-3}}\right)^{-1} \rm yr,\label{eq:tgas}
\ena 
which is long compared to the spin-up time (Eq. \ref{eq:tspinup}). Therefore, for supersonic drift, grains can be spun-up to suprathermal rotation before the damping by gas collisions.

Rapidly spinning nanoparticles emit strong electric dipole radiation, which can also damp grain rotation on timescale of
\bea
\tau_{\rm ed}=\frac{3I^{2}c^{3}}{\mu^{2}kT_{\gas}}\simeq  225 \left(\frac{a_{-7}^{7}}{3.8\hat{\beta}}\right)\left(\frac{1000\K}{ T_{\gas}}\right)\yr,\label{eq:taued}
\ena
where $\mu$ is the dipole moment and $\hat{\beta}=\beta/(0.4\D)$ with $\beta$ being the dipole moment per structure of the grain (see Section \ref{sec:spindust}), and the small contribution of dipole moment due to asymmetric charge distribution on the grain surface is ignored for numerical convenience (\citealt{1998ApJ...508..157D}; \citealt{Hoang:2010jy}). 

Comparing $\tau_{\rm ed}$ with $\tau_{\H}$, one can see that, for hot and dense shocked regions, the electric dipole damping rate is lower than the gas damping rate. Thus, we expect the excitation by gas collisions and neutral drift is very efficient in spinning nanoparticles up to extremely fast rotation because electric dipole emission is not fast enough to remove the grain angular momentum.

\subsection{Grain charge distribution}
{In the shock code used for computing the shock structure, calculations of grain charge distribution are not yet included. Therefore, to find the grain charge distribution,} we take the physical parameters of the shock as input parameters for our charging code (\citealt{2011ApJ...741...87H}; \citealt{2012ApJ...761...96H}).

{The temperatures $T_{e},T_{i},T_{n},T_{d}$, number density of neutrals, ion species and ionization fraction ($x_{\H}$ and $x_{M}$)} at each location in the shock obtained from the shock model are used to find the collisional charging rate. Photoelectric emission is calculated as in \cite{2001ApJS..134..263W} (see also \citealt{2012ApJ...761...96H}), assuming the similar radiation spectrum as the ISRF, but with the strength $\chi$ shown in Table \ref{tab:ISM}.\footnote{Note that the collisional charging time is comparable to the mean time between two collisions with electrons is $1/(n_{e}v\pi a^{2})\sim 1\rm yr$, which is much shorter than the flow time. Thus, we can find the charge distribution of nanoparticles in C-shocks.}

\begin{figure}
\includegraphics[width=0.45\textwidth]{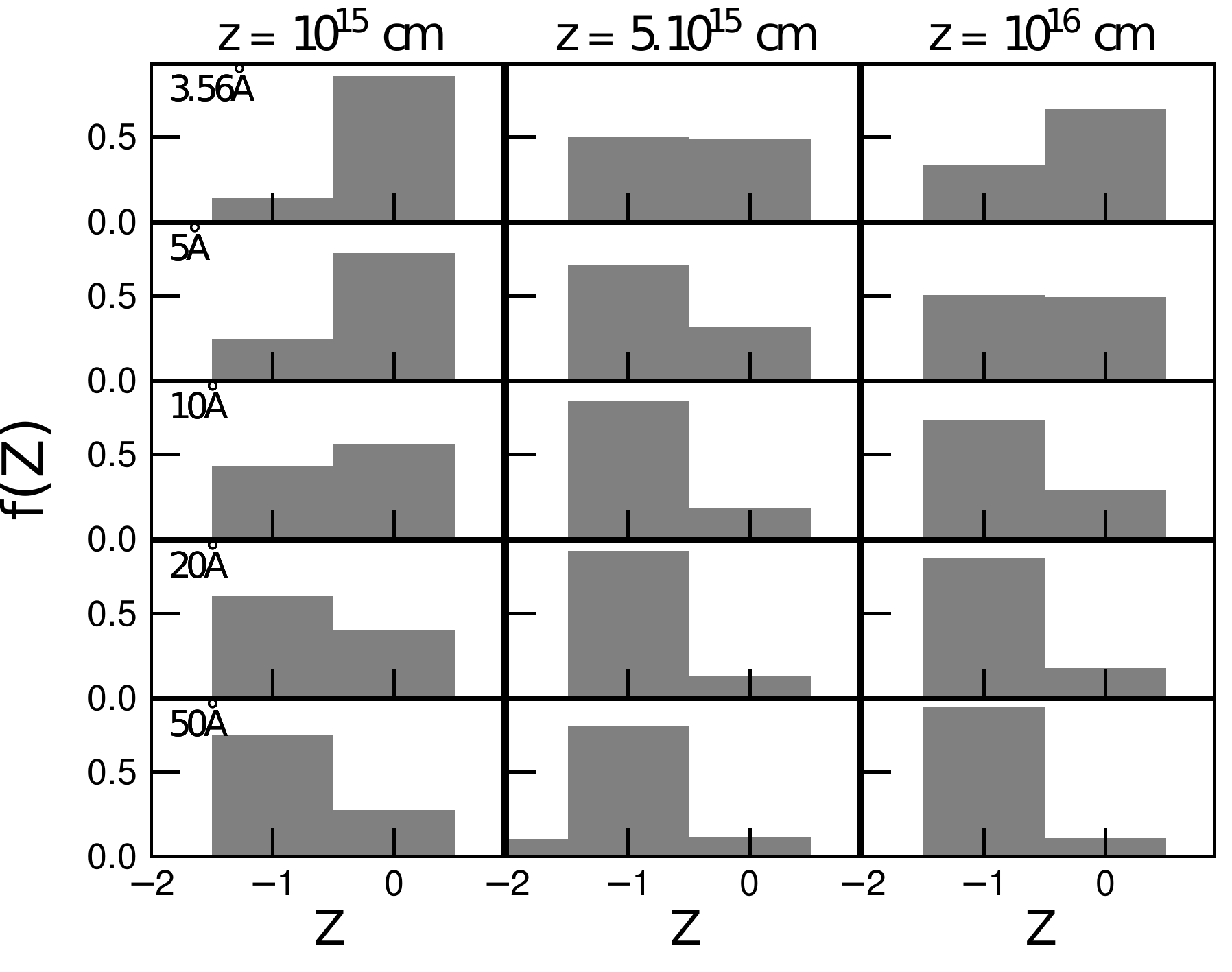}
\caption{Charge distribution functions for nanoparticles of radii a = 3.56, 5, 10, 20, and 50\AA~ at three positions in the shock with $n_{\H}=10^{4}\cm^{-3}$ and $v_{\s}=10\km\s^{-1}$.}
\label{fig:charge}
\end{figure}

Figure \ref{fig:charge} shows the charge distribution of PAHs at the different locations in the shock. For grains below $10$\AA, the grain charge is mostly between $Z=-1$ and $Z=0$, with a considerable probability on $Z=-1$. The fraction of grains being on the negative charge states is increasing with increasing $a$, and it also varies with the distance in the shock. Since the neutral gas-grain drift is only relevant for charged grains, a high probability of being on $Z=-1$ indicates that grain excitation by neutral drift is important.

\subsection{Numerical results for damping and excitation coefficients}

To account for the charge distribution of nanoparticles, we can write the net damping and excitation coefficients due to grain-neutral collisions as follows:
\bea
F_{n}=F_{n,s_{d}=0}f_{Z}(Z=0) + \sum_{Z\ne 0}F_{n,s_{d}\ne 0}f_{Z}(Z),\\
G_{n}=G_{n,s_{d}=0}f_{Z}(Z=0) + \sum_{Z\ne 0}G_{n,s_{d}\ne 0}f_{Z}(Z),
\ena
where the first terms for $s_{d}=0$ are calculated as in \cite{1998ApJ...508..157D}, and the second terms describe the effect of the charged grain drift in the shock which are given by Equation (\ref{eq:Gsd}).

\begin{figure}
\includegraphics[width=0.45\textwidth]{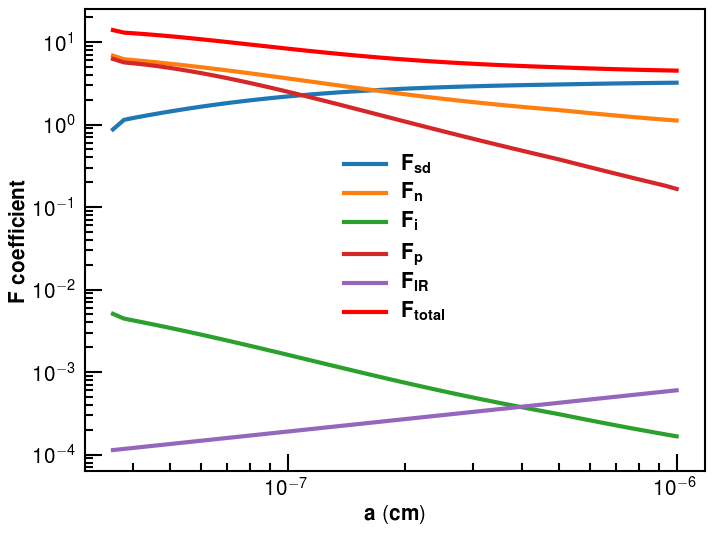}
\includegraphics[width=0.45\textwidth]{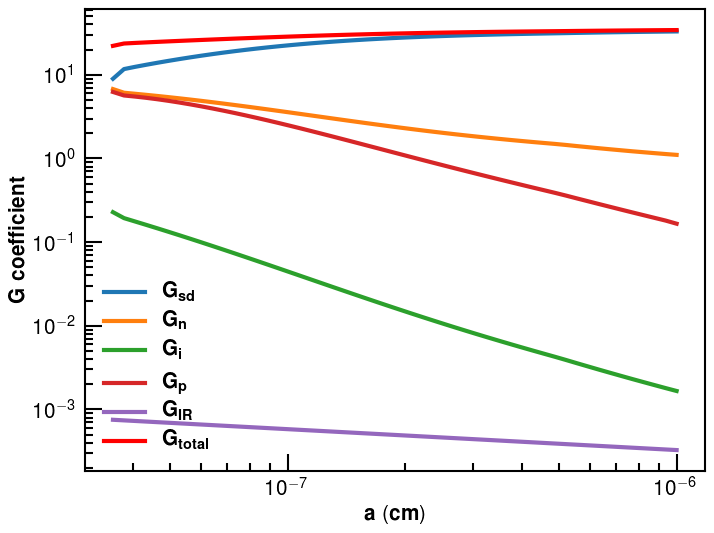}
\caption{Rotational damping and excitation coefficients from various interaction processes in the C-shock of velocity $s_{d}=4$ computed at location $z=5.54\times 10^{14}\cm$. Excitation by neutral drift is dominant. The pre-shock gas density $n_{\H}=10^{4}\cm^{-3}$ is considered.}
\label{fig:FG_coeff}
\end{figure}

Figure \ref{fig:FG_coeff} shows the damping (upper panel) and excitation (lower panel) coefficients from the various interaction processes, assuming a supersonic drift with $s_{d}=4$. The rotational damping is dominated by neutral impact and plasma drag for small nanoparticles. As expected, rotational excitation by neutral-grain drift is dominant, while the excitation by infrared emission is negligible due to low radiation intensity. Interestingly, $F_{sd}$ and $G_{sd}$ tend to increase with increasing grain size $a$, while the coefficients of the other processes decrease. This can be understood in terms of grains having an increased fraction of the negative charge states (see Figure \ref{fig:charge}).

\subsection{Rotational temperature and Rotation Rate}
Let $T_{\rm rot}$ be the rotational temperature of spinning nanoparticles, so that $I\langle\omega^{2}\rangle/2=3kT_{\rm rot}/2$. Thus, using the rms angular velocity from \cite{1998ApJ...508..157D}, we obtain
\bea
\frac{T_{\rm rot}}{T_{\rm gas}}=\frac{G}{F}\frac{2}{1 + [1+ (G/F^2)(20\tH/3\ted)]^{1/2}},\label{eq:Trot}
\ena
where $\tH$ and $\ted$ are the characteristic damping times due to gas collisions and electric dipole emission. For $n_{\H}\ge 10^{4}\cm^{-3}$, one can see from Equations (\ref{eq:tgas}) and (\ref{eq:taued}) that $\tau_{ed}\gg \tau_{\H}$.

\begin{figure}
\includegraphics[width=0.45\textwidth]{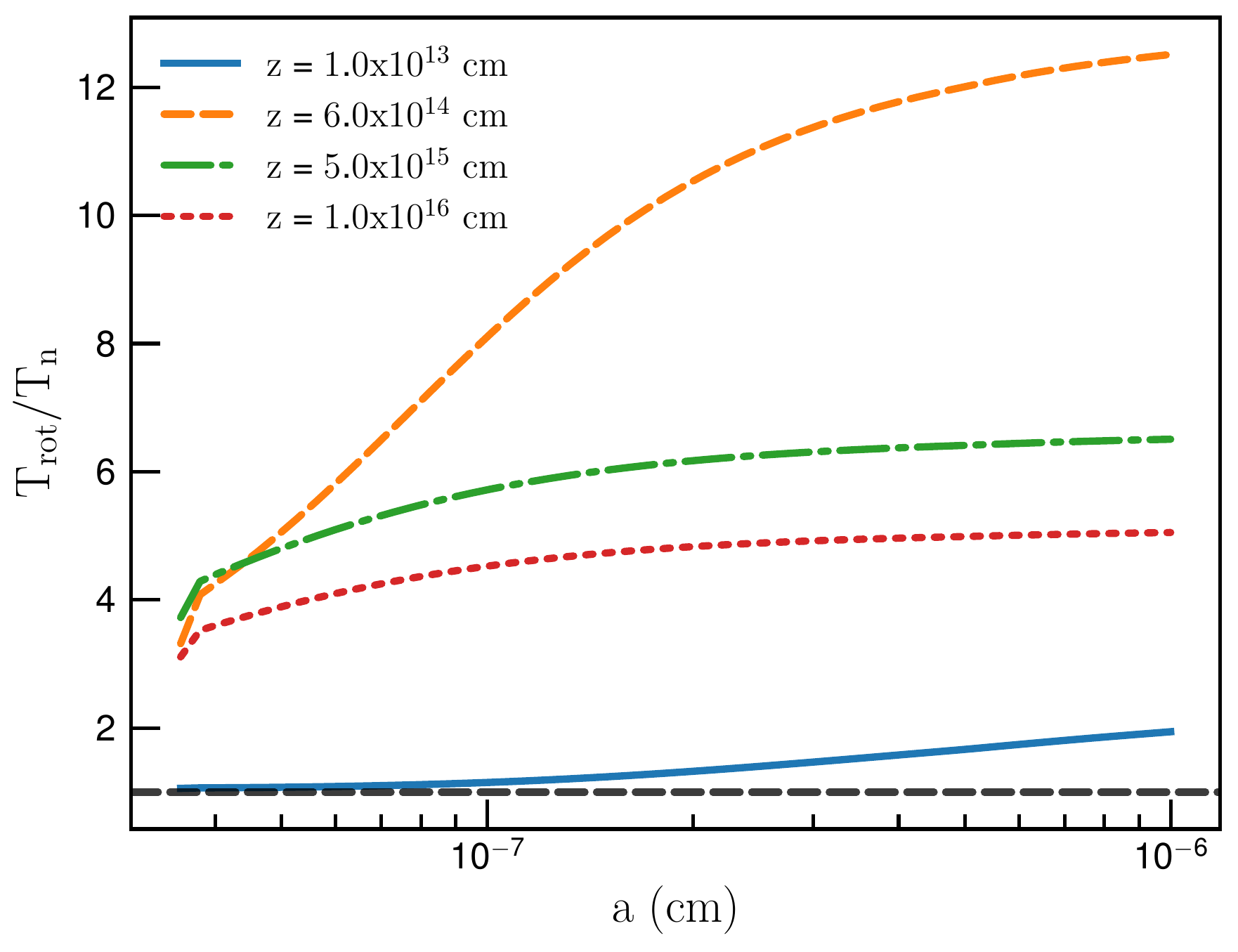}
\includegraphics[width=0.45\textwidth]{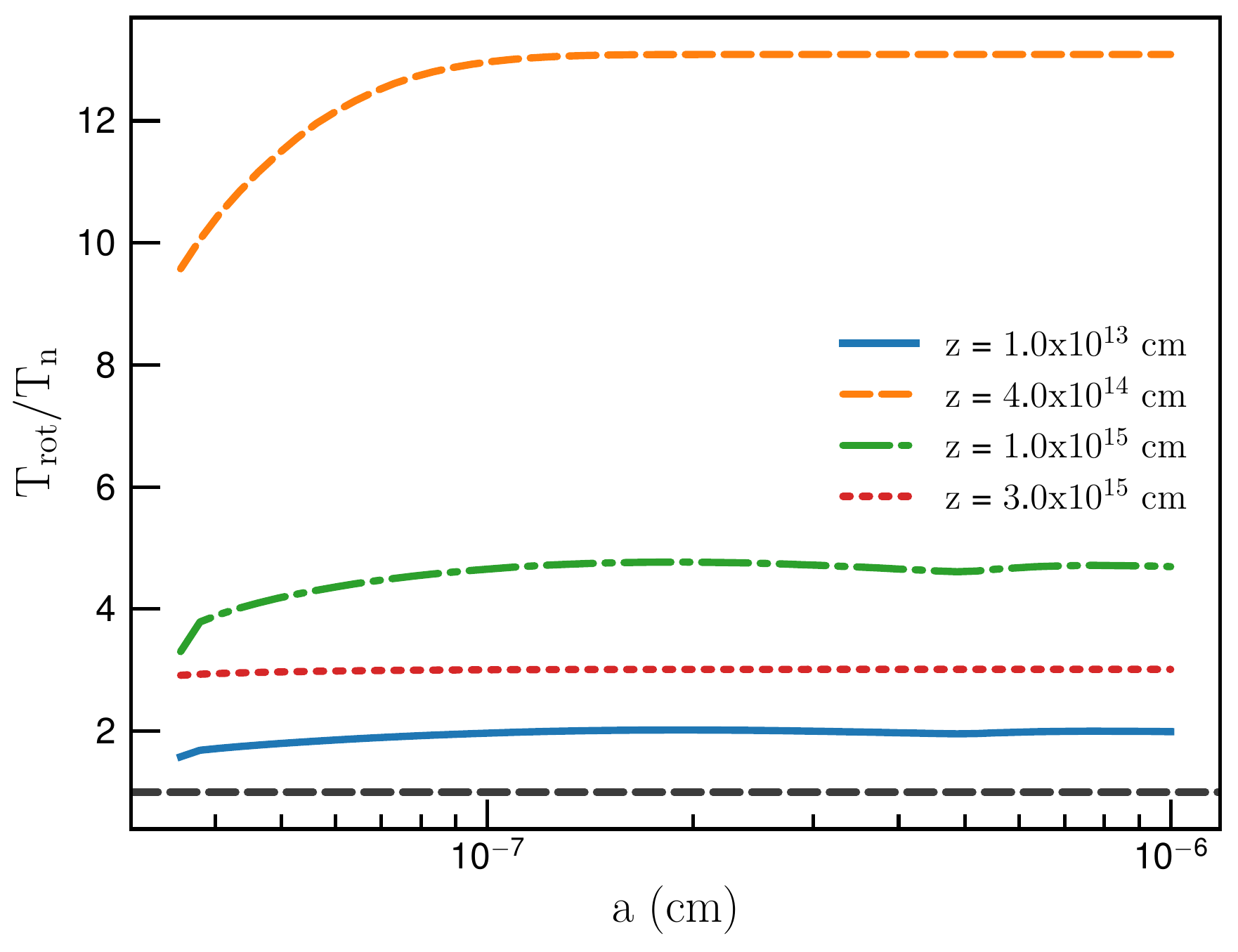}
\includegraphics[width=0.45\textwidth]{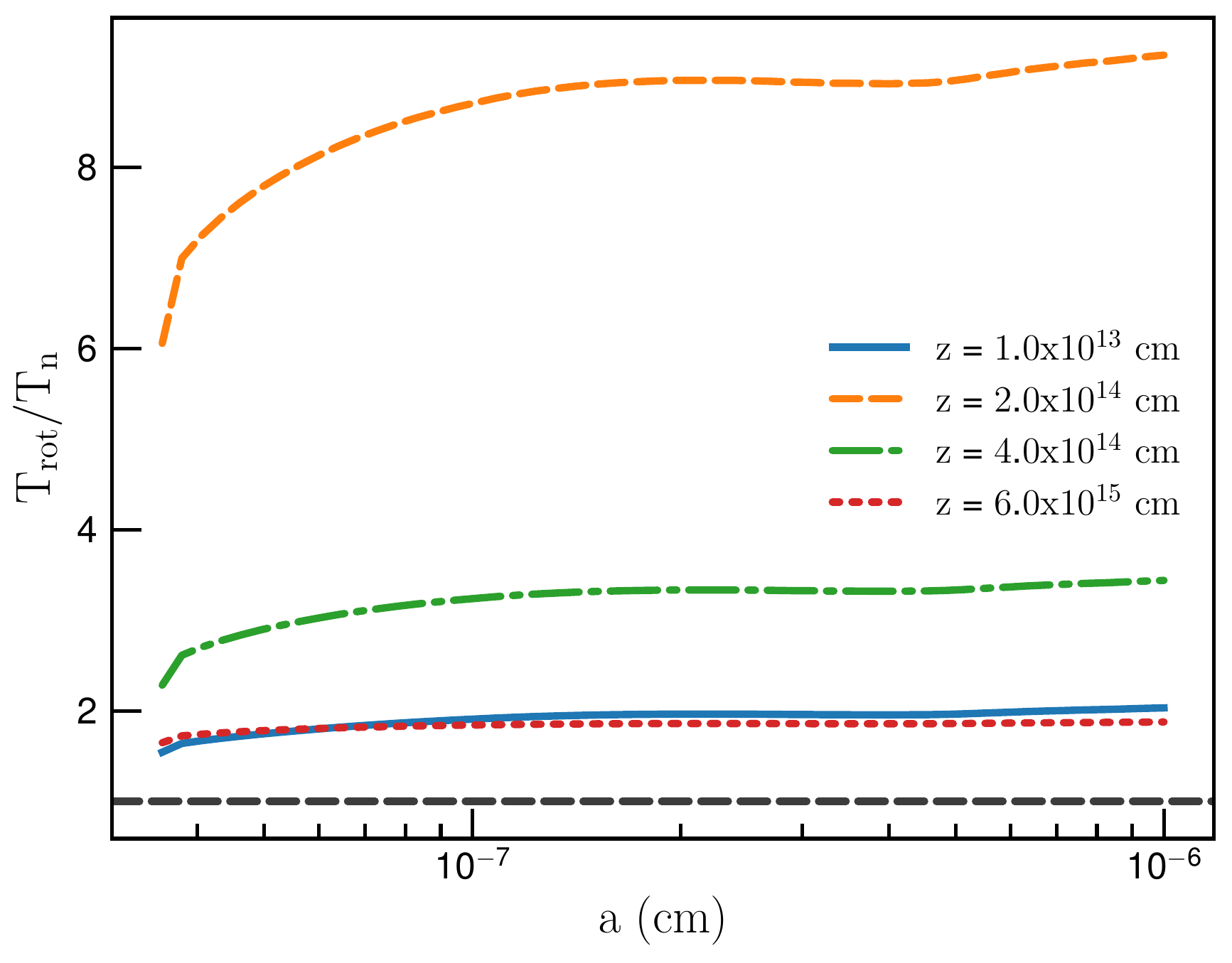}
\caption{Variation of $T_{\rm rot}/T_{\rm n}$ with grain size $a$ for $v_{s}=20 \km\s^{-1}$ and $n_{\H}=10^{4}\cm^{-3}$ (top panel- model A), $n_{\H}=10^{5}\cm^{-3}$ (middle panel- model B) and $n_{\H}=10^{6}\cm^{-3}$ (bottom panel- model C). Suprathermal rotation is observed at most locations considered except the early stage with $z=10^{13}\cm$ (blue line) in the case of $n_{\H}=10^{4}\cm^{-3}$.}
\label{fig:Trot_a}
\end{figure}

Figure \ref{fig:Trot_a} shows the rotational temperature relative to the neutral gas temperature, $T_{\rm rot}/T_{\gas}$, as a function of grain size, at four different locations in the shock for three shock models. Nanoparticles can indeed rotate suprathermally due to supersonic neutral drift. The ratio $T_{\rm rot}/T_{n}$ increases with grain size due to an increasing probability of being on the negative charge states. The ratio $T_{\rm rot}/T_{n}$ also varies with the shock location and is larger inside the shock. 

The rotation rate at the rotational temperature $T_{\rm rot}$ is given by
\bea
\frac{\omega_{\rm rot}}{2\pi}&=&\frac{1}{2\pi}\left(\frac{3kT_{\rm rot}}{I}\right)^{1/2}\nonumber\\
&\simeq& 1.4\times 10^{10}\hat{\rho}^{-1/2}a_{-7}^{-5/2}\left(\frac{T_{\rm rot}}{10^{3}\K}\right)^{1/2}\Hz,\label{eq:omega_Trot}
\ena
where $\hat{\rho}=\rho/3\g\cm^{-3}$. Thus, for $T_{\rm rot}\gtrsim 10^{3}\K$, nanoparticles of $a=1$ nm can rotate extremely fast at $\omega_{\rm rot}\gtrsim 10^{10} \s^{-1}$.

\section{Rotational disruption for extremely fast rotating nanoparticles}\label{sec:disrupt} \label{sec:rotational_disruption}
\subsection{Rotational disruption}
In this section, we introduce a new mechanism of dust destruction in shocks which is based centrifugal stress within rapidly spinning nanoparticles, namely rotational disruption. 

The basic idea of rotational disruption is as follows. A spherical dust grain rotating at velocity $\omega$ develops a centrifugal stress due to centrifugal force, which scales as $S=\rho a^{2} \omega^{2}/4$ (\citealt{2018arXiv181005557H}). When the rotation rate increases to a critical limit such that the tensile stress induced by centrifugal force exceeds the maximum tensile stress, so-called tensile strength of the material, nanoparticles are disrupted instantaneously. The critical angular velocity for the disruption is given by
\bea
\frac{\omega_{\rm cri}}{2\pi}&=&\frac{1}{\pi a}\left(\frac{S_{\rm max}}{\rho} \right)^{1/2}\nonumber\\
&\simeq& 1.8\times 10^{11}a_{-7}^{-1}\hat{\rho}^{-1/2}S_{\rm max,10}^{1/2}\Hz,~~~~\label{eq:omega_cri}
\ena
where $S_{\rm max}$ is the tensile strength of dust material and $S_{\rm max,10}=S_{\rm max}/10^{10}\erg\cm^{-3}$ is the tensile strength in units of $10^{10}\erg\cm^{-3}$.\footnote{An alternative unit of the tensile strength is ${\rm dyne/cm^{2}}$, but in this paper we use the unit of $\erg\cm^{-3}$ for $S_{\rm max}$.}

The exact value of $S_{\max}$ depends on the dust grain composition and structure. Compact grains can have higher $S_{\max}$ than porous/composite grains. Ideal material without impurity, such as diamond, can have $S_{\max}\ge 10^{11}\erg\cm^{-3}$ (see \cite{2018arXiv181005557H} for more details). In the following, nanoparticles with $S_{\max} \gtrsim 10^{10}\erg\cm^{-3}$ are referred to as strong materials, and those with $S_{\max} < 10^{10}\erg\cm^{-3}$ are called weak materials. 

\begin{figure}
\includegraphics[width=0.45\textwidth]{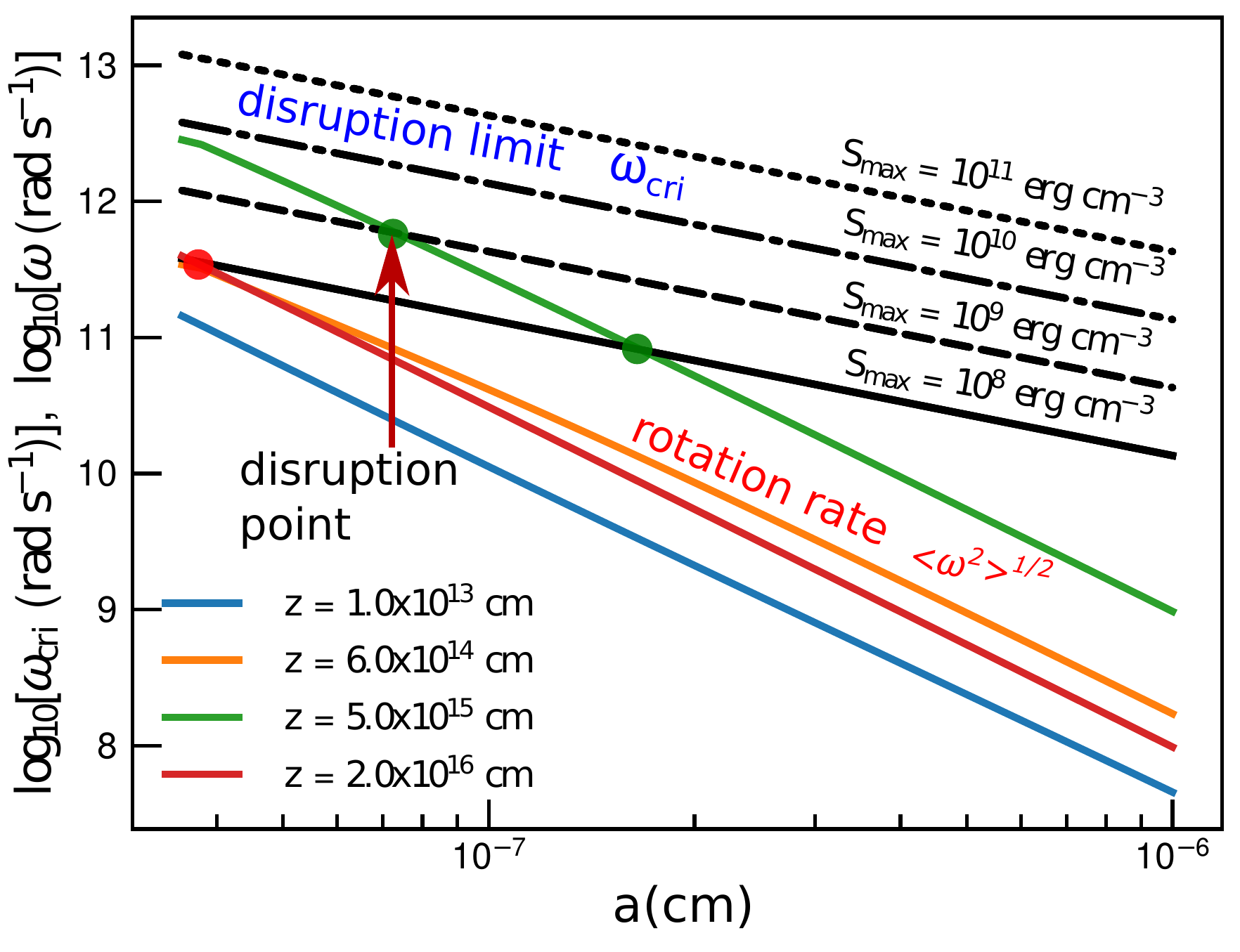}
\caption{Angular rotational velocity of nanoparticles vs. grain size in comparison with critical velocity of rotational disruption $\omega_{\rm cri}$ for several shocked positions. Smallest nanoparticles are disrupted, while larger ones can survive. Shock density $n_{\H}=10^{4}\,$cm$^{-3}$ and $v_{s}=20\km\s^{-1}$ are considered. Filled circles mark the disruption location.}
\label{fig:omega_cri}
\end{figure}

Figure \ref{fig:omega_cri} shows the rms rotation rate $\langle\omega^{2}\rangle^{1/2}$ as a function of grain size. The critical disruption rate for the different tensile strengths is also shown for comparison. For instance, nanoparticles of size $a< 0.8$ nm (marked by red arrow) can be disrupted by the centrifugal force at $S_{\max}=10^{9}\erg\cm^{-3}$ with shock density $n_{\H}=10^{4}\,$cm$^{-3}$ and shock velocity $v_{s}=20\,$km$\,$s$^{-1}$.

\subsection{Disruption rotational temperature}
From Equations (\ref{eq:omega_Trot}) and (\ref{eq:omega_cri}) one can infer the rotational temperature required for grain disruption:
\bea
T_{\rm rot}&\ge& \left(\frac{32\pi a^{3}}{45k}\right)S_{\max}\nonumber\\
&\simeq&1.6\times10^{5}a_{-7}^{3}S_{\rm max,10}\K,
\ena
which follows that $T_{\rm rot}\sim 20000\K$ is required to destroy nanoparticles of $a\le 0.5$ nm, assuming the typical $S_{\max}=10^{10}\erg\cm^{-3}$. 

For a lower tensile strength of $S_{\max}=10^{9}\erg\cm^{-3}$, the disruption can occur at temperature $T_{\rm rot}\sim 2000\K$ for $a\le 0.5$ nm, i.e., smallest nanoparticles can be disrupted by thermal gas collisions without needing supersonic neutral drift.

\subsection{Disruption time and critical drift velocity}
Using the excitation coefficients obtained in Section \ref{sec:times}, we can evaluate the time required to spin-up nanoparticles to $\omega_{\rm cri}$, so-called rotational disruption time, as follows:
\bea
\tau_{\rm disr}&=&
\frac{J_{\rm cri}^{2}}{(\Delta J)^{2}/(\Delta t)}=\frac{(I\omega_{\rm cri})^{2}}{n_{\H}m_{\H}^{2}v_{\rm drift}^{3}\pi a^{4}}\label{eq:tdisr}\\
&\simeq&0.05 a_{-7}^{4}n_{5}^{-1}S_{\rm max,10}\left(\frac{v_{\rm drift}}{10\km\s^{-1}}\right)^{-3}~{\rm yr},\nonumber~~~
\ena
where $n_{5}=n_{\H}/(10^{5}\cm^{-3})$. 

The disruption time is much lower than the flow time $t_{\rm flow}$ for a typical drift velocity of $v=10\km\s^{-1}$. Thus, rotational disruption is important in shocked regions. Moreover, the disruption time scales as $a^{4}$, thus, small nanoparticles tend to be disrupted faster than large grains. 

For shock velocities of $v_{s}<50\km\s^{-1}$, the gas temperature is not high enough such that thermal rotation can disrupt nanoparticles. As a result, to disrupt grains, one requires $t_{\rm disr}<\tau_{\H}$, which corresponds to
\bea
v_{\rm drift}\ge 9a_{-7}\left(\frac{T_{\gas}}{1000\K}\right)^{1/6}S_{\rm max,10}^{1/3}\km\s^{-1}.\label{eq:vcri}
\ena

From Figures (\ref{fig:Cshockn4-velo}) and (\ref{fig:Cshockn6-velo}) one can see that the drift velocity in C-shocks can easily satisfy the above condition.

\begin{figure}
\includegraphics[width=0.45\textwidth]{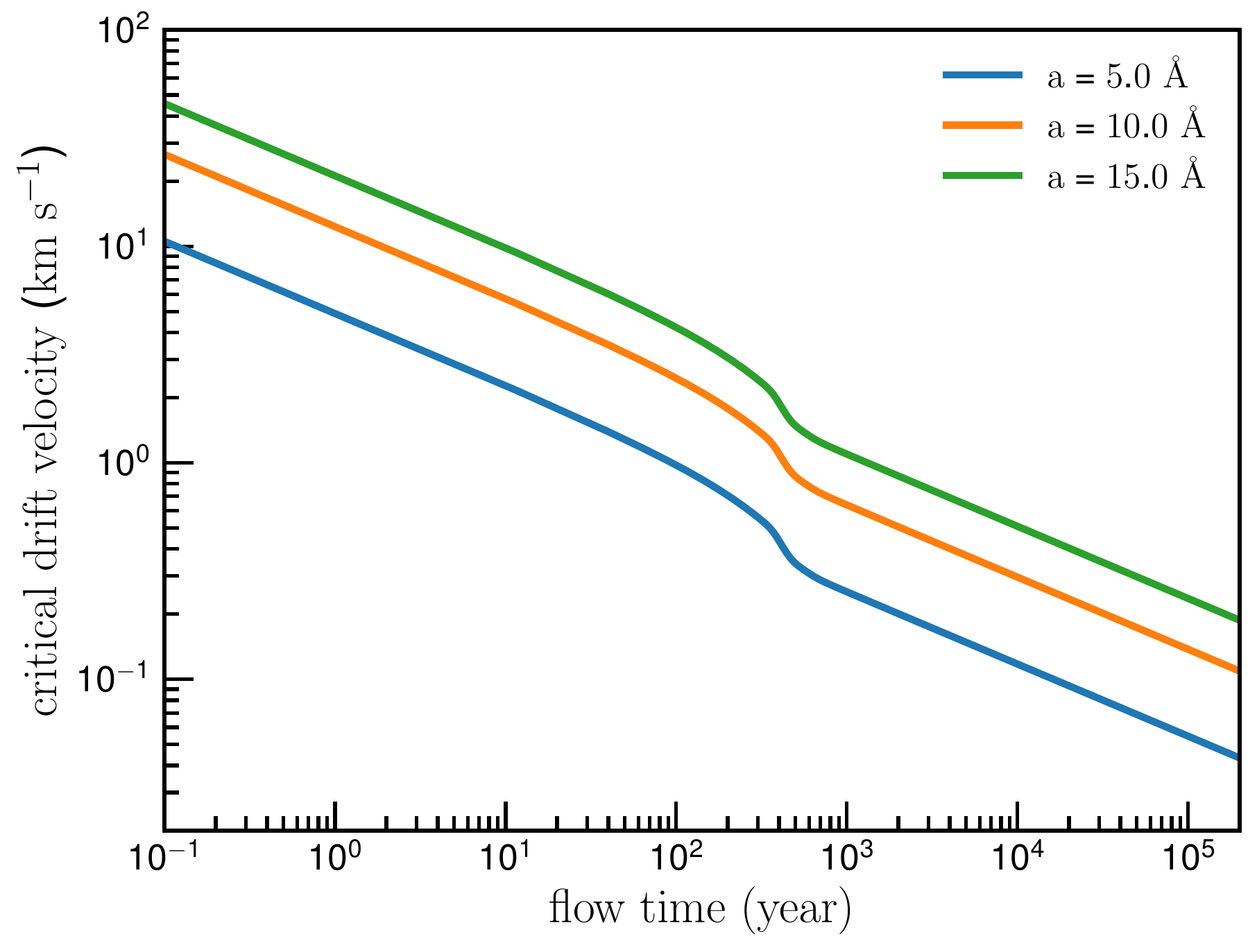}
\caption{Critical drift velocity of which nanoparticles are disrupted vs. flow time for three grain sizes, assuming $n_{\H}=10^{4}\cm^{-3}$ and $v_{\s}=20\km\s^{-1}$. $S_{\max}=10^{10}\erg\cm^{-3}$ is considered.}
\label{fig:vdrift_cri}
\end{figure}

Figure \ref{fig:vdrift_cri} shows the critical drift velocity above which nanoparticles are disrupted, as a function of flow time. Several grain sizes and tensile strength are considered. The critical velocity decreases rapidly with increasing flow time.

\subsection{Disruption grain size}

To calculate the smallest size $a_{\rm min}$ that nanoparticles can withstand the rotational disruption, we compute $\langle \omega^{2}\rangle$ using the rotational temperature $T_{\rm rot}$ as given by Equation (\ref{eq:Trot}) at each shock location for a grid of grain sizes from $0.35-10$ nm and compare it with $\omega_{\rm cri}$.

\begin{figure}
\includegraphics[width=0.45\textwidth]{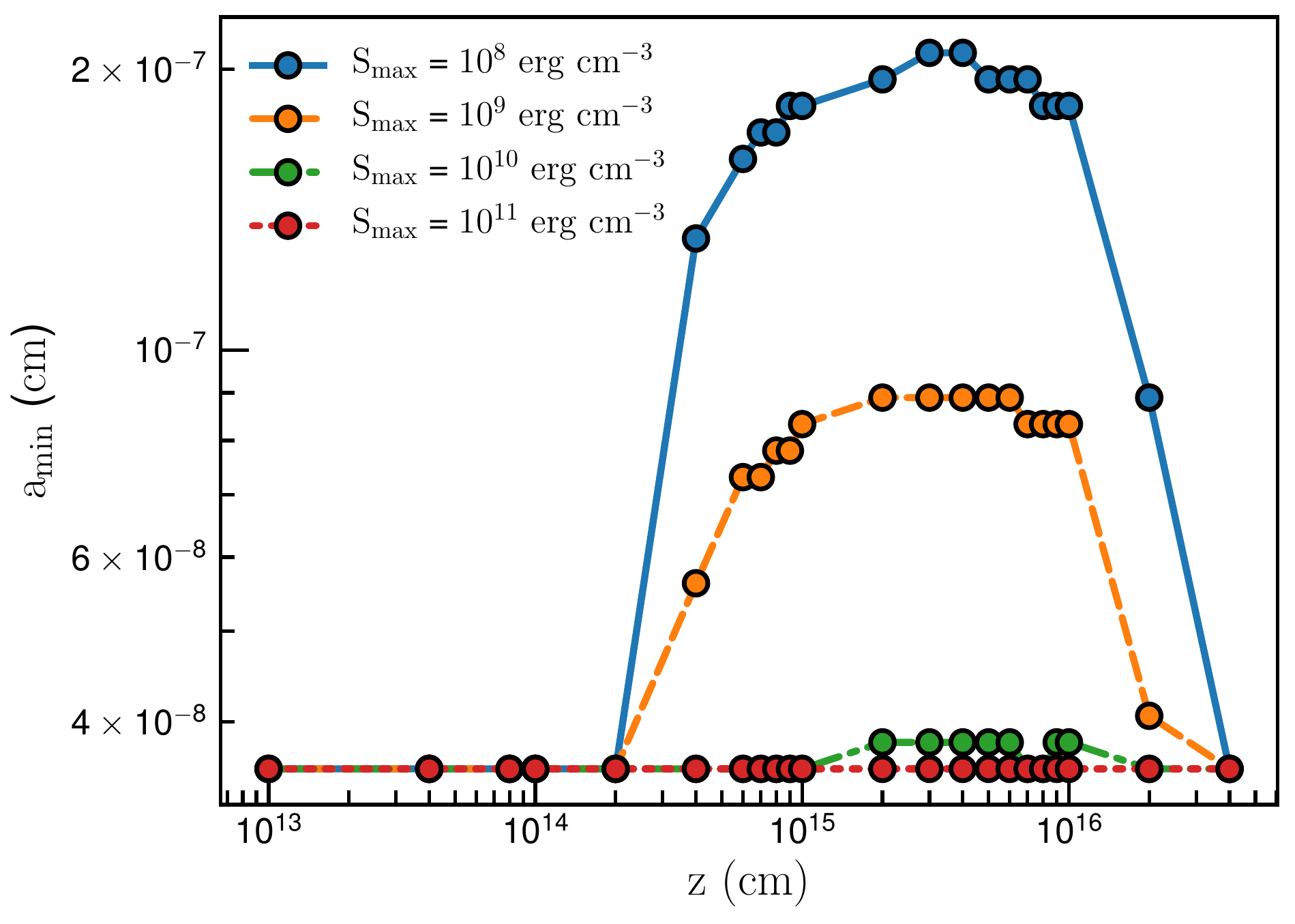}
\includegraphics[width=0.45\textwidth]{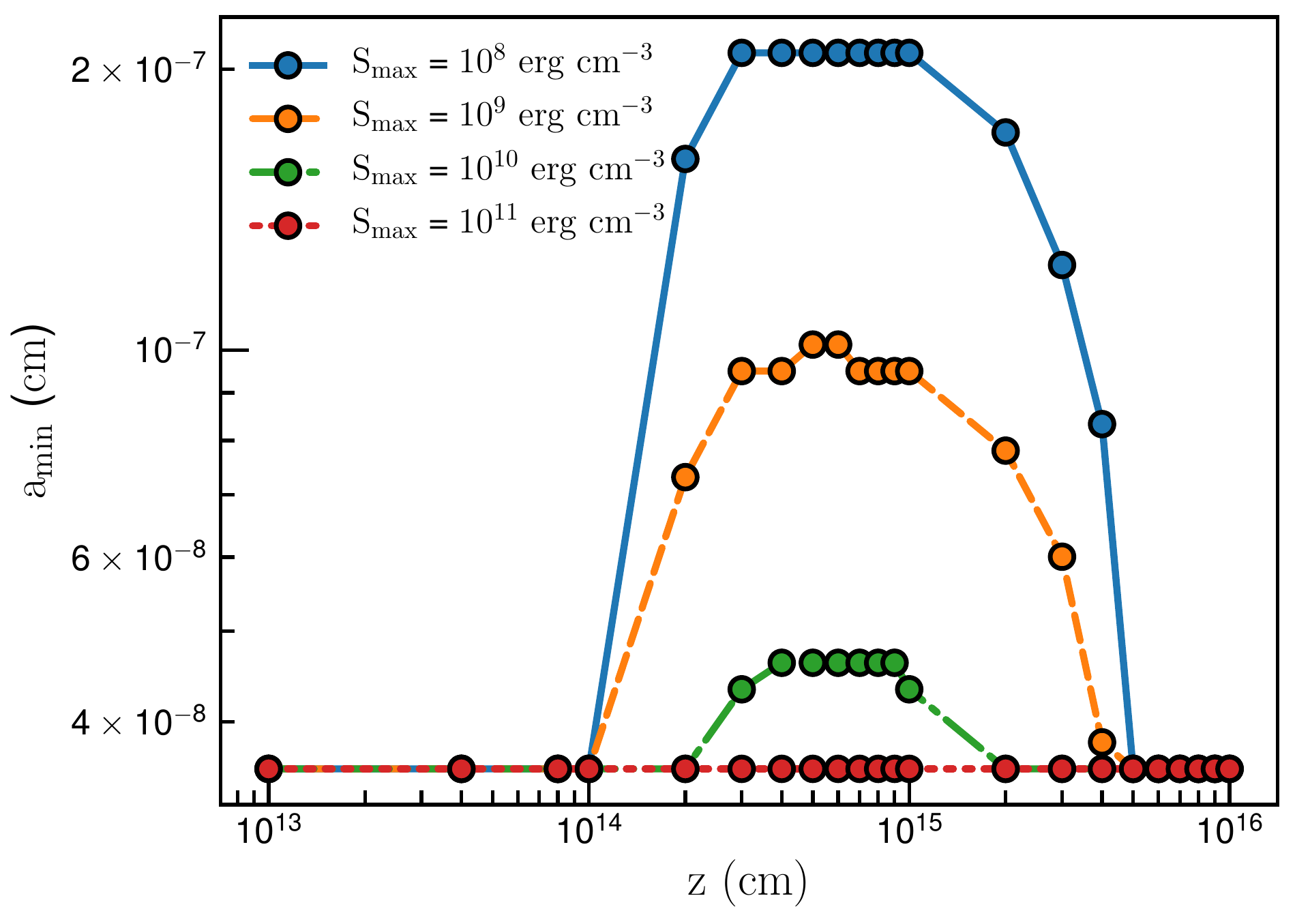}
\includegraphics[width=0.45\textwidth]{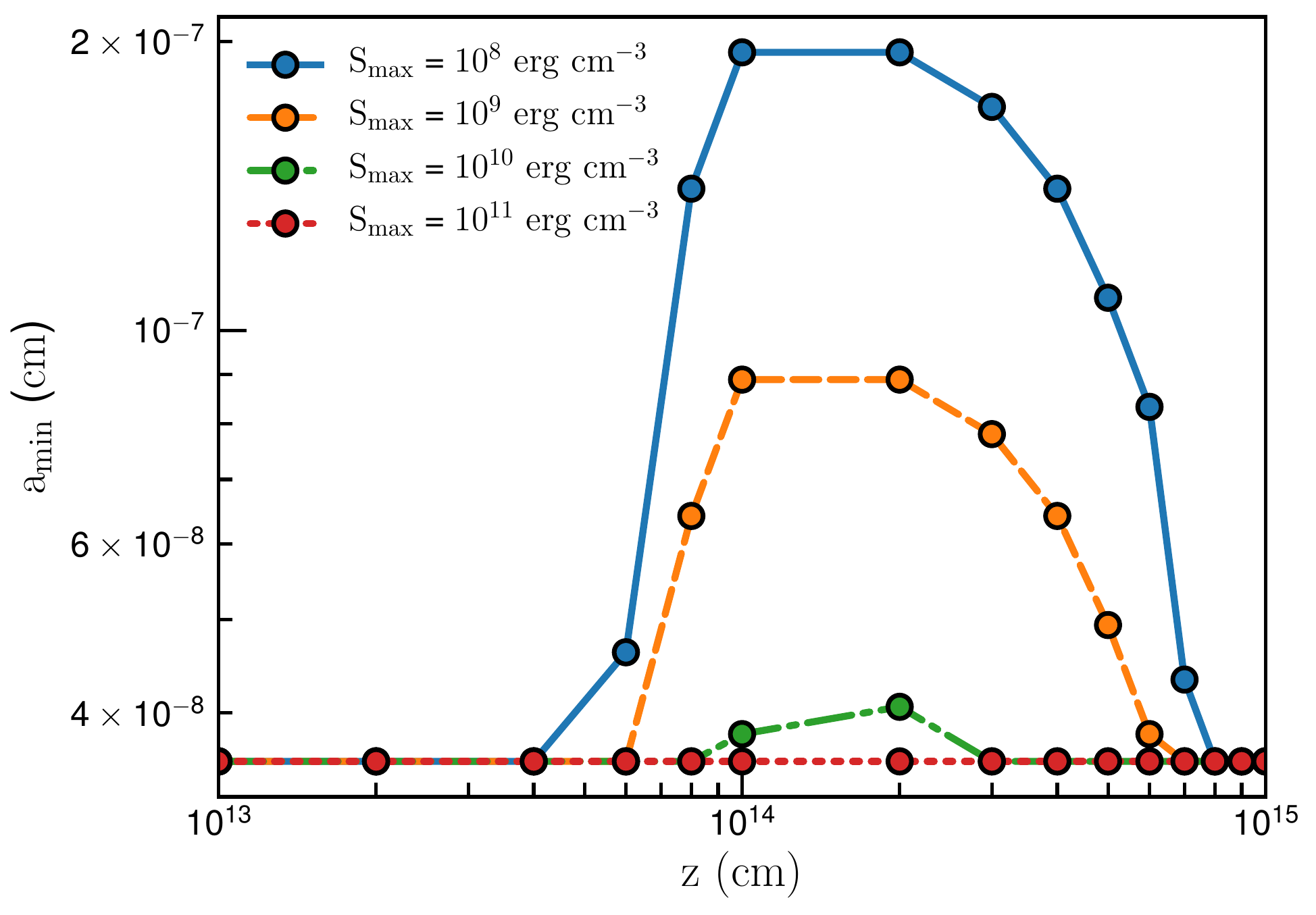}
\caption{Minimum size below which PAHs are destroyed by rotational disruption vs. distance in the shock, assuming the different material tensile strengths for $v_s=30\km\s^{-1}$. Three shock models with gas density $n_{\H}=10^{4}\cm^{-3}$ (top panel, model A),
$n_{\H}=10^{5}\cm^{-3}$ (middle panel, model B), and
$n_{\H}=10^{6}\cm^{-3}$ (bottom panel, model C).}
\label{fig:a_cri}
\end{figure}

Figure \ref{fig:a_cri} shows the obtained minimum size $a_{\rm min}$ as a function of distance in the shock for different values of $S_{\rm max}$ and three shock models. Strong nanoparticles can survive the shock passage (red line), while weak nanoparticles can destroyed. { Grain disruption size increases toward the middle of the shock and then rapidly declines, which resembles the temperature and velocity profile of shocks (see e.g., Figure \ref{fig:Cshock-temp}).} The disruption size is below 0.5 nm for the typical $S_{\rm max,10}=1$, but it can be increased to $2.0$ nm for weaker materials (see blue, orange and green lines). The disruption size is largest for the shock model B (middle panel).

\section{Spinning dust emission from nanoparticles in C-shocks}\label{sec:spindust}
To pave the way for future observations, in this section, we will calculate spinning dust emission from nanoparticles in shocked regions, accounting for rotational disruption.

\subsection{Spinning dust model}
\subsubsection{Electric dipole moment and emission power}
The rotational emission mechanism is built upon the assumption that nanoparticles own non-zero electric dipole moments. PAH molecules can acquire intrinsic dipole moments due to polar bonds (see \citealt{1998ApJ...508..157D}). The attachment of SiO and SiC molecules to the grain surface gives rise to the electric dipole moment for nanosilicates (\citealt{2016ApJ...824...18H}). 

Let $N$ be the total number of atoms in a spherical nanoparticle of radius $a$. Assuming PAHs with a typical structure C:H=$3:1$ having mean mass per atom $m\approx 9.25$ amu, one obtains $N=545a_{-7}^{3}$ for the mass density $\rho=2\g\cm^{-3}$ (\citealt{1998ApJ...508..157D}). Assuming nanosilicates with structure SiO$_{4}$Mg$_{1.1}$Fe$_{0.9}$ having $m=24.15$ amu, one has $N=418a_{-7}^{3}$ for $\rho=4\g\cm^{-3}$ \citep{2016ApJ...824...18H}.

Let $\beta$ be the dipole moment per atom in the grain. Assuming that dipoles have a random orientation distribution, the intrinsic dipole moment of the grain can be estimated using the random walk formula:
\bea
\mu^{2}=N\beta^{2}\simeq 86.5(\beta/0.4\D)^{2} a_{-7}^{3} \D^{2},\label{eq:muin}
\ena
for PAHs, and $\mu^{2}\simeq 66.8(\beta/0.4\D)^{2} a_{-7}^{3} \D^{2}$ for nanosilicates \citep{2016ApJ...824...18H}.

The power emitted by a rotating dipole moment $\mu$ at angular velocity $\omega$ is given by the Larmor formula:
\bea
P(\omega,\mu)=\frac{2}{3}\frac{\omega^4\mu^2\sin^2\theta}{c^3}~~~,
\ena
where $\theta$ is the angle between $\bomega$ and $\bmu$. Assuming an uniform distribution of the dipole orientation, $\theta$, then, $\sin^{2}\theta$ is replaced by $\langle \sin^{2}\theta\rangle=2/3$.

\subsubsection{Angular momentum distribution function}
In dense regions where gas-grain collisions dominate rotation dynamics of nanoparticles (e.g., in shocked regions), the grain angular velocity can be appropriately described by the Maxwellian distribution: 
\bea
f_{\rm MW}(\omega, T_{\rm rot})=\frac{4\pi}{ (2\pi)^{3/2}}\frac{I^{3/2}\omega^{2}}{(kT_{\rm rot})^{3/2}}\exp\left(-\frac{I\omega^{2}}{2kT_{\rm rot}} \right),\label{eq:fomega}
\ena
where $I$ is the moment of inertia of the spherical nanoparticle of mass density $\rho$, and $T_{\rm rot}$ is the grain rotational temperature.

\subsubsection{Size distribution: PAHs and nanosilicates}

Following \cite{Li:2001p4761}, nanoparticles are assumed to follow a log-normal size distribution:
\bea
\frac{1}{n_{\H}}\frac{dn_{j}}{da} = \frac{B_{j}}{a}\exp\left(-0.5\left[\frac{\log (a/a_{0,j})}{\sigma_{j}}\right]^{2}\right) ,\label{eq:dnda_log}
\ena
where $j=PAH, sil$ corresponds to PAHs and nanosilicate composition, $a_{0,j}$ and $\sigma_{j}$ are the model parameters, and $B_{j}$ is a constant determined by
\bea
B_{j}&=&\frac{3}{(2\pi)^{3/2}}\frac{{\rm exp}(-4.5\sigma_{j}^{2})}{\rho \sigma a_{0,j}^{3}}\nonumber\\
&&\times \left(\frac{m_{X}b_{X}}{1+{\rm erf}[3\sigma/\sqrt{2} + {\rm ln}(a_{0}/a_{\min})/\sigma\sqrt{2}}\right),\label{eq:Bconst}
\ena
where $m_{X}$ is the grain mass per atom X, $b_{X}=X_{\H}Y_{X}$ with $Y_{X}$ being the fraction of $X$ abundance contained in very small sizes and $X_H$ being the solar abundance of element $X$. In our studies, $X=$C for PAHs and $X=$Si for nanosilicates. In addition, $m_{X}=m_{C}$ for PAHs, and $m_{X}=m(SiO_{4}Mg_{1.1}Fe_{0.9})$ for nanosilicates of the adopted composition. 

The peak of the mass distribution $a^{3}dn_{j}/d\ln a$ occurs at $a_{p}=a_{0,j}e^{3\sigma_{j}^{2}}$. Three parameters determine the size distribution of nanoparticles, including $a_{0,j},\sigma_{j}, Y_{X}$. 

\subsubsection{Spinning dust emissivity and emission spectrum}
Let $j_{\nu}^{a}(\mu, T_{\rm rot})$ be the emissivity from a spinning nanoparticle of size $a$ at a location in the shock, where $T_{\rm rot}$ depends on local conditions in the shock (see Section \ref{sec:rot}). Thus,
\bea
j_{\nu}^{a}(\mu, T_{\rm rot})= \frac{1}{4\pi}P(\omega,\mu) {\rm pdf}(\nu|\omega)=\frac{1}{4\pi}P(\omega,\mu)2\pi f_{\rm MW}(\omega),\label{eq:jem_a}
\ena
where $pdf(\nu|\omega)$ is the probability that the nanoparticle rotating at $\omega$ emits photons at observe frequency $\nu$, and the relation $\omega=2\pi \nu$ is assumed.

{Here we disregarded the effect of grain wobbling \citep{Hoang:2010jy} and assume that nanoparticles are rotating along one axis as in \citealt{1998ApJ...508..157D}. This assumption is likely appropriate for shocked regions because suprathermal rotation ($T_{\rm rot}\gtrsim T_{\rm gas}\gg T_{d}$) due to supersonic neutral drift is expected to induce rapid alignment of the axis of maximum inertia moment with the angular momentum (i.e., internal alignment, \citealt{1979ApJ...231..404P}).}

The rotational emissivity per H nucleon is obtained by integrating over the grain size distribution (see \citealt{2011ApJ...741...87H}):
\bea \label{eq:jnu_w}
\frac{j_{\nu}(\mu, T_{\rm rot})}{n_{\H}}=\int_{a_{\min}}^{a_{\max}}j_{\nu}^{a}(\mu,T_{\rm rot})\frac{1}{n_{\H}} \frac{dn}{da} da,\label{eq:jem}
\ena 
where $dn/da = dn_{\rm PAH, sil}/da$ for spinning PAHs and nanosilicates, respectively.

The minimum size $a_{\rm min}$ at each location $z$ in the C-shock is set equal to the disruption size (see Figure \ref{fig:a_cri}), which is a function of the shock velocity. 

\subsection{Spinning Dust Emissivity}
We use Equation (\ref{eq:jnu_w}) to calculate the spinning dust emissivity at various locations inside the shock. The emissivity is calculated assuming that dust composes of 90 $\%$ PAHs and 10 $\%$ nanosilicates. In the absence of rotation disruption, a$_{\rm min}$ is taken to be equal to 3.56 $\AA$. When the rotational disruption effect is taken into account, a$_{\rm min}$ is determined as in Section \ref{sec:rotational_disruption}. We fix the abundance of PAHs and nanosilicates throughout the shock, although their abundance should vary in the shock due to grain shattering (\citealt{2011A&A...527A.123G}).

\begin{figure}
\includegraphics[width=0.45\textwidth]{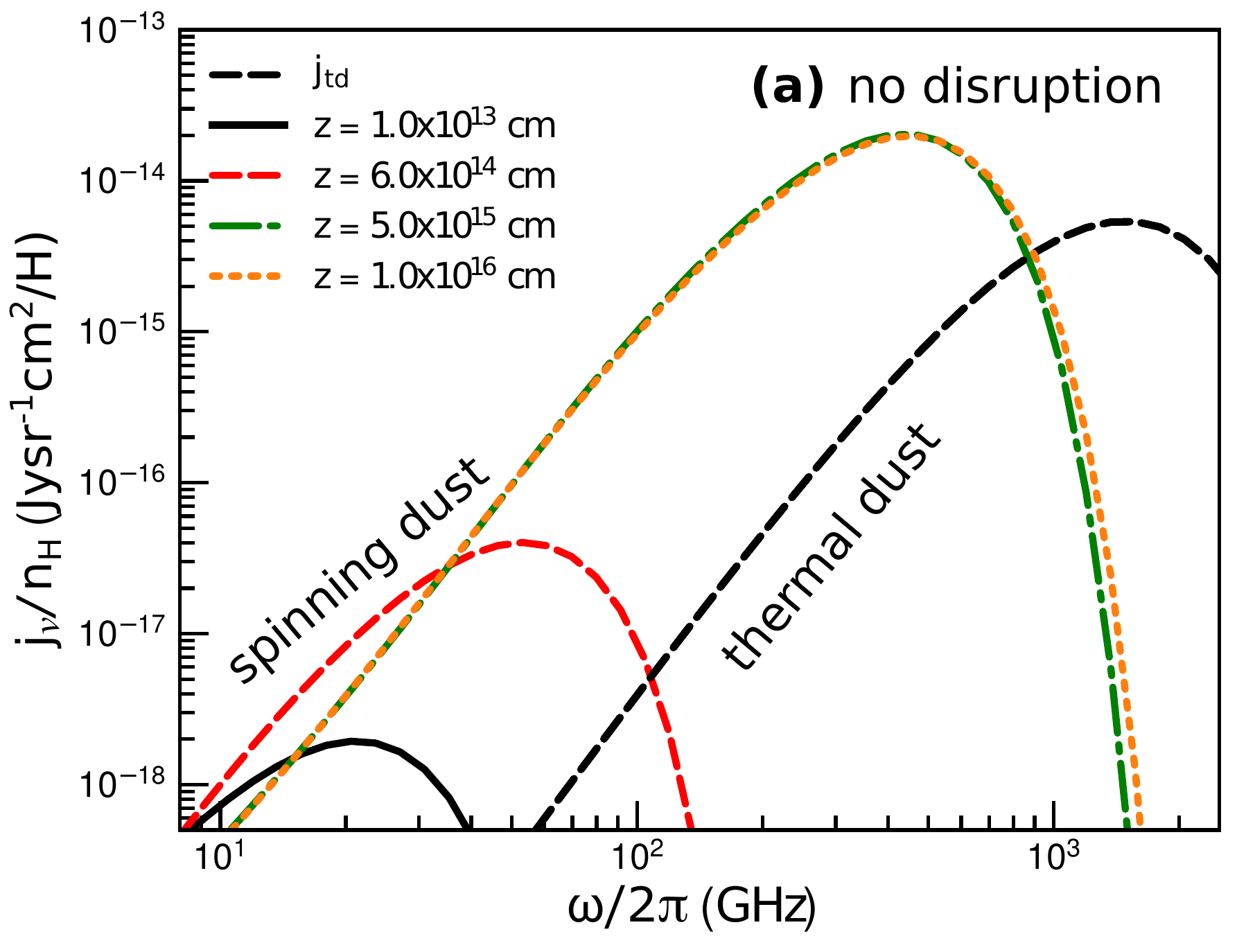}
\includegraphics[width=0.45\textwidth]{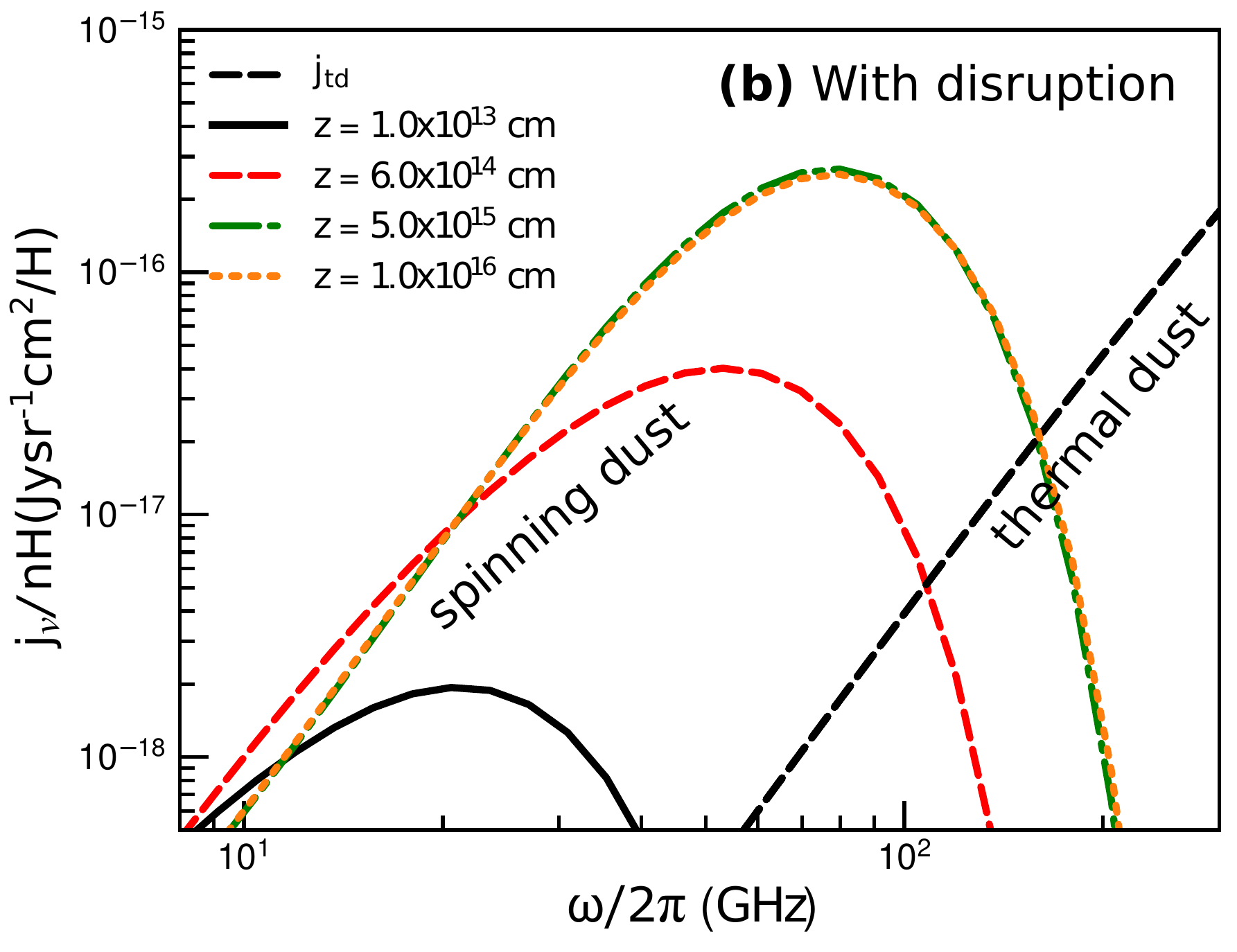}
\caption{Rotational emission spectrum from spinning nanoparticles for $n_{\H}=10^{4}\cm^{-3}$ and $v_{s}=20 \km\s^{-1}$ computed at several positions in the shock without rotational disruption (panel (a)) and with rotational disruption for $S_{\rm max}=10^{10}\erg\cm^{-3}$ (panel (b)). Dust is considered of $90\%$ of PAH and $10\%$ of silicate. Thermal dust emission from large grains is shown in dashed black line for comparison.}
\label{fig:spindust_e}
\end{figure}

Figure \ref{fig:spindust_e} shows spinning dust emissivity from nanoparticles computed at different location $z$ in the shock. The black dashed line shows thermal dust emission from large grains (see \citealt{Hoang:2018hc}). {When the rotational disruption effect is not taken into account, spinning dust emissivity is very strong and can peak at very high frequencies of $\nu\sim 500$ GHz for some locations (see panel (a)). When accounting for rotational disruption (panel (b)), both rotational emissivity and peak frequency are reduced significantly due to the destruction of the smallest nanoparticles via rotational disruption.} The effect of rotational disruption is clearly demonstrated through emission spectrum at locations $z=5\times 10^{15}\cm$ and $z=10^{16}\cm$, where the peak frequency is reduced from $\nu\sim 500$ GHz (panel (a)) to $\nu\sim 80$ GHz (panel (b)). In both cases, spinning dust is still dominant over thermal dust at frequencies $\nu<100$ GHz (lower panel).

\subsection{Emission spectral flux}
The spectral flux of spinning dust emission from nanoparticles in a spherical shocked region can be calculated as follows:
\bea \label{eq:flux}
F_{\nu}=\frac{L_{\nu}}{4\pi D^{2}}&=&\frac{1}{4\pi D^{2}}\int_{V} n_{\H}dV \left(\frac{4\pi j_{\nu}}{n_{\H}}\right) \nonumber \\
&=&\frac{1}{4\pi D^{2}}\int_{z_{i}}^{z_{f}}4\pi z^{2}dz n_{\H}\left(\frac{4\pi j_{\nu}}{n_{\H}}\right),
\ena
where $D$ is the distance from the shocked region to the observer, $V$ is the shock volume, and $z_{i},z_{f}$ are the initial and final distances of the shock. {Here $z_{i}$= 10$^{13}\,$cm, and $z_{f}$ is also the shock length, determined by the shock distance where $T_{\gas}$ decreases to the pre-shocked gas temperature.}

\begin{figure}
\includegraphics[width=0.45\textwidth]{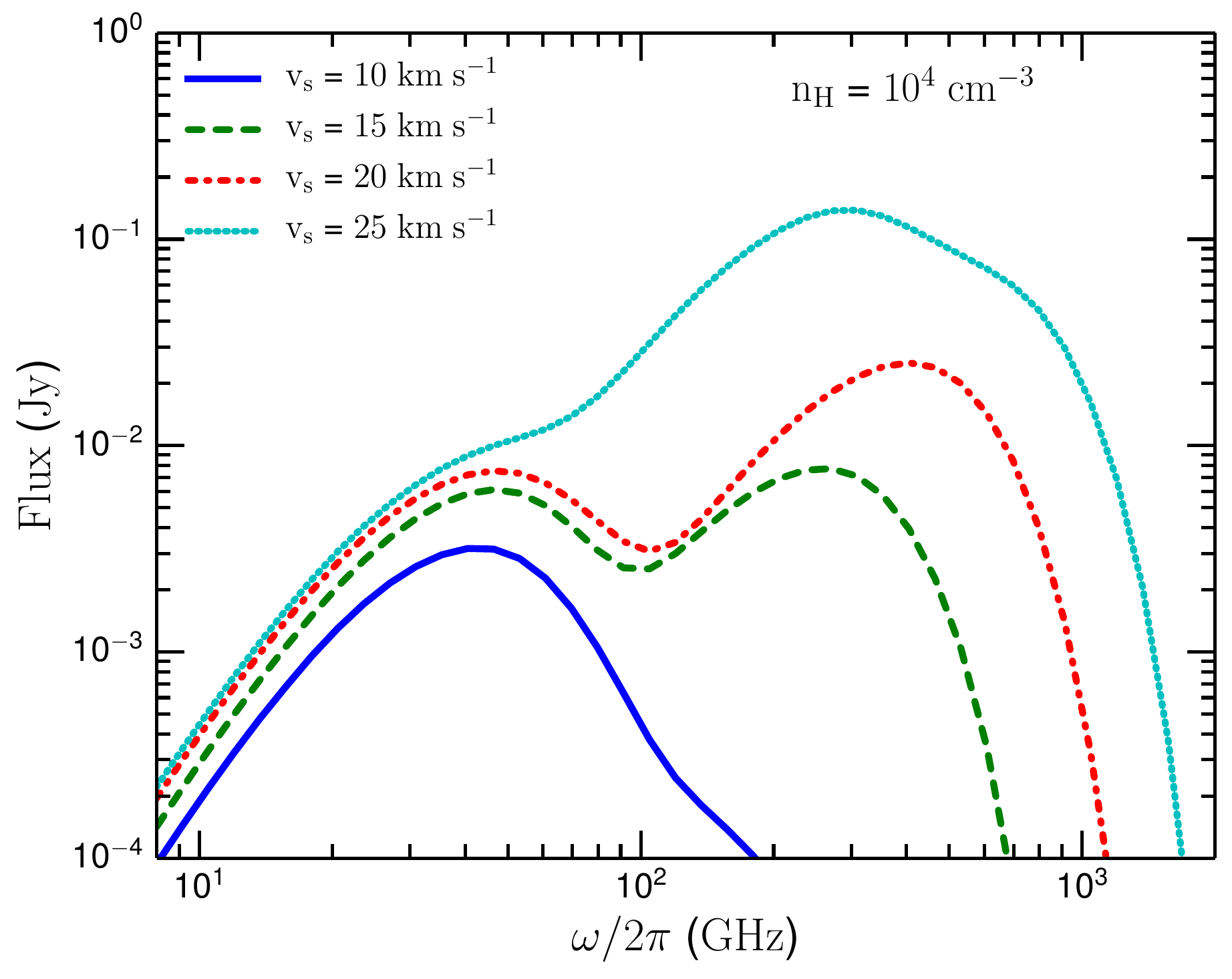}
\includegraphics[width=0.45\textwidth]{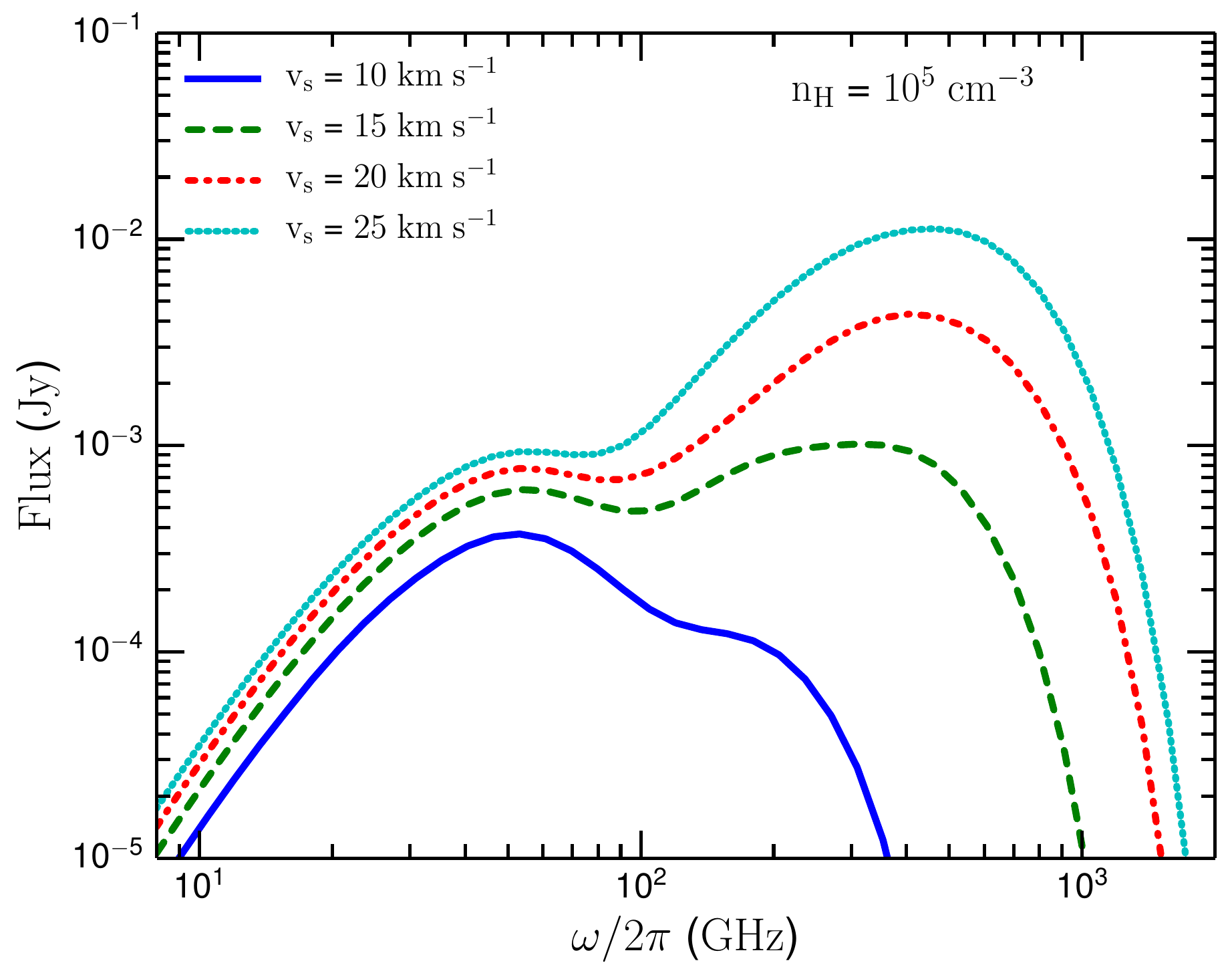}
\includegraphics[width=0.45\textwidth]{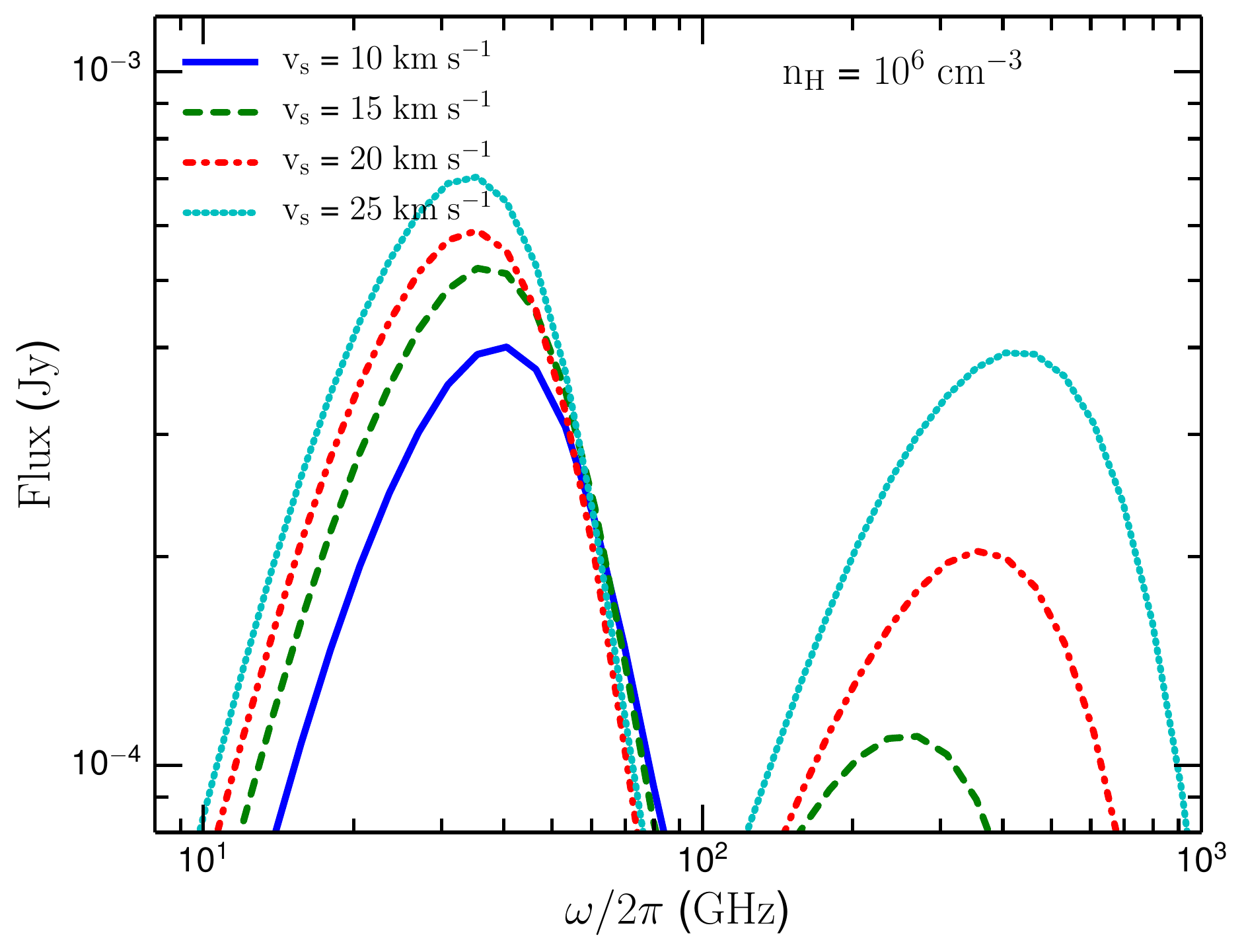}
\caption{Spectral flux of spinning dust emission for three shock models with $n_{\H}=10^{4}\cm^{-3}$ (top panel, model A), $n_{\H}=10^{5}\cm^{-3}$ (middle panel, model B), and $n_{\H}=10^{6}\cm^{-3}$ (bottom panel, model C). Emission spectrum has two peaks, a low-frequency and a high-frequency one. The value $S_{\max}=10^{10}\erg\cm^{-3}$, and D=100 pc are considered.}
\label{fig:flux_S10}
\end{figure}

{Figure \ref{fig:flux_S10} shows the spectral flux for three shock models of different velocities, assuming nanoparticles made of strong material ($S_{\rm max}=10^{10}\erg\cm^{-3}$). For this case, rotational disruption is quite inefficient for this case (see Figure \ref{fig:a_cri}). One interesting feature is that the emission spectrum consists of two parts. The first, low-frequency ($\nu<100$ GHz) part with peak flux at frequency $\nu \sim 40$ GHz, and the second, high-frequency ($\nu>100$ GHz) part with peak flux at $\nu\sim 300-500$ GHz. The second part is produced by smallest nanoparticles which are excited to suprathermal rotation by supersonic neutral drift (see Figure \ref{fig:spindust_e} for emissivity). Furthermore, the peak flux decreases with increasing $n_{\H}$ due to the decreased shock length $z_{f}$ (see Figures \ref{fig:Cshockn4-velo} and \ref{fig:Cshockn6-velo}).}

\begin{figure}
\includegraphics[width=0.45\textwidth]{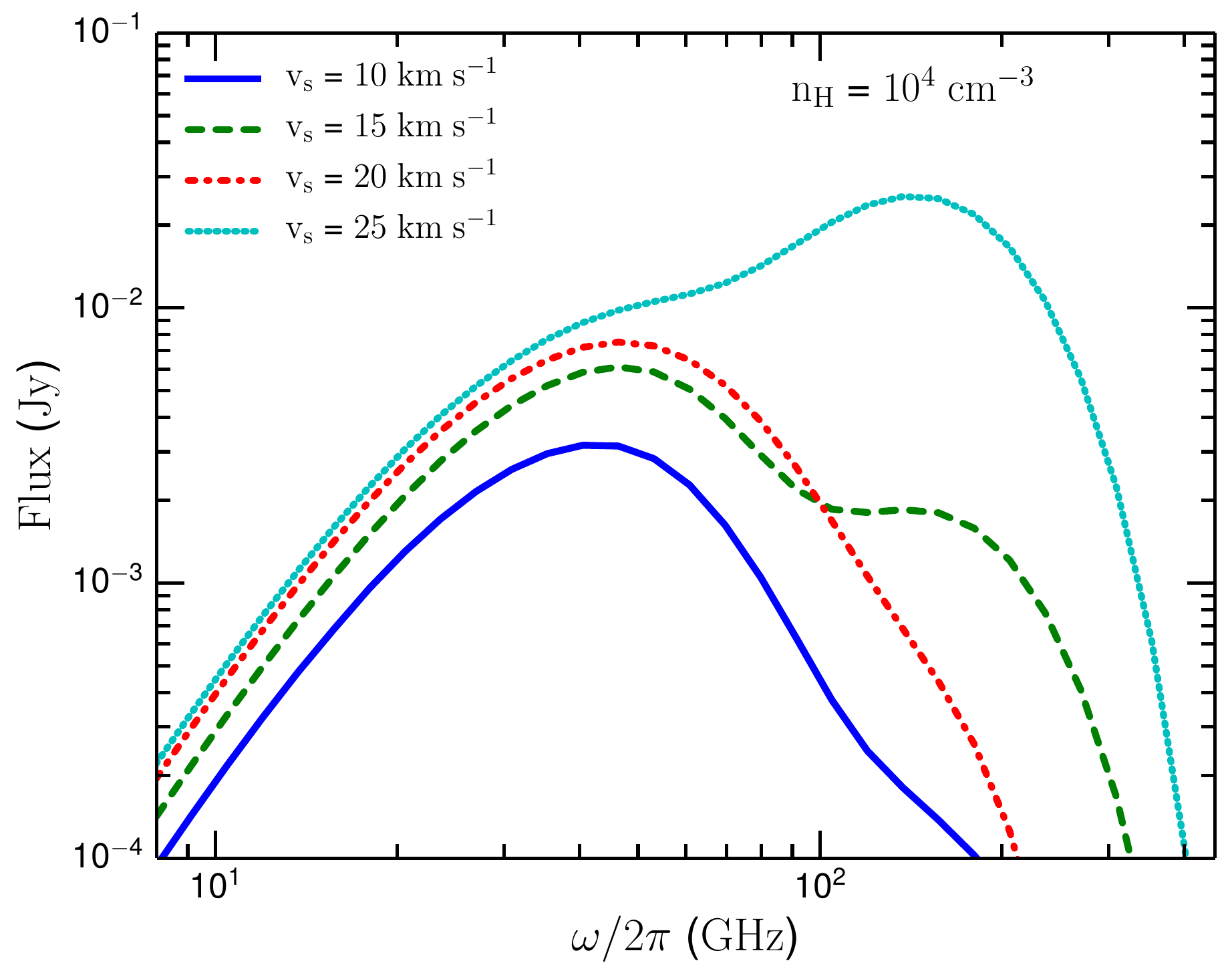}
\includegraphics[width=0.45\textwidth]{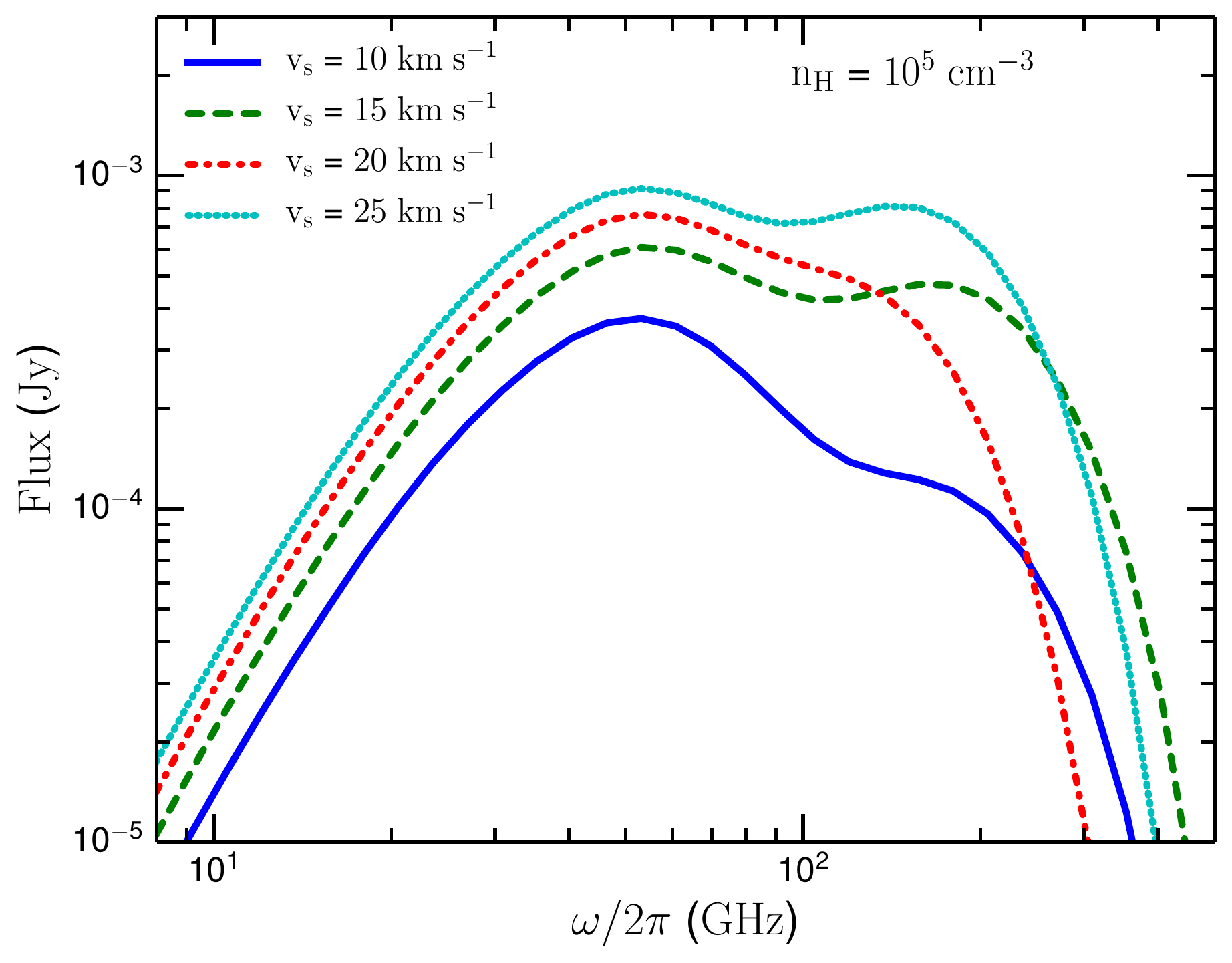}
\includegraphics[width=0.45\textwidth]{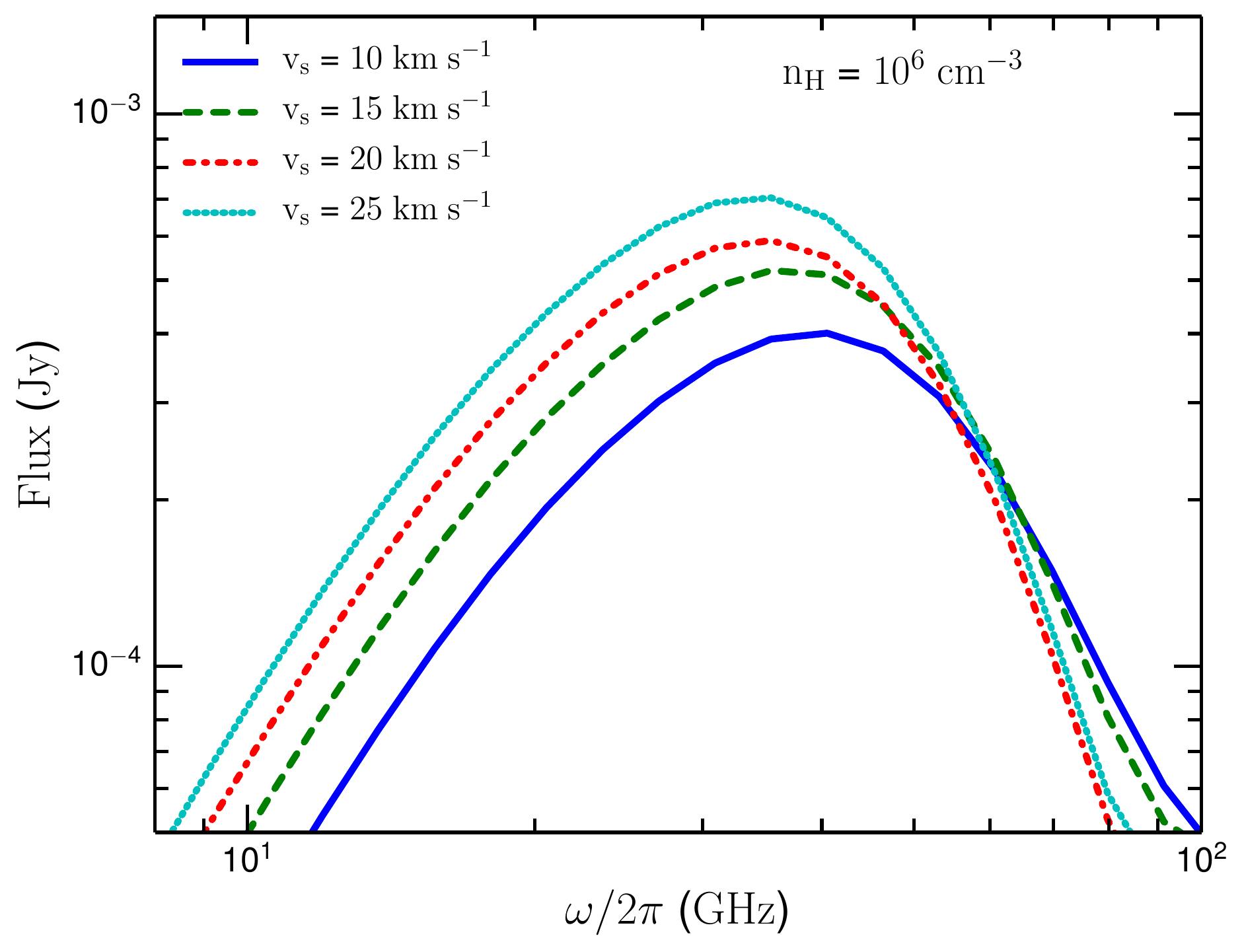}
\caption{Same as Figure \ref{fig:flux_S9} but for weaker nanoparticles of $S_{\max}=10^{9}\erg\cm^{-3}$. The high-frequency peak is mostly destroyed due to the removal of small nanoparticles by rotational disruption.}
\label{fig:flux_S9}
\end{figure}

{Figure \ref{fig:flux_S9} shows similar results but for nanoparticles made of weaker materials such that rotational disruption is more efficient. Compared to Figure \ref{fig:flux_S10}, one can see that the emission spectrum is similar to the first part of Figure \ref{fig:flux_S10}, but the high-frequency peak is almost wiped out. The reason is that the smallest nanoparticles are significantly reduced due to efficient rotational disruption, resulting in the suppression of the high-frequency part in the emission spectrum.}

\section{Discussion}\label{sec:discuss}

\subsection{Dust processing in shocks: comparisons of rotational disruption with other mechanisms}
In shocked regions, shattering of large grains due to collisions between neutral and charged grains is suggested to produce PAHs and silicate nanoparticles, resulting in the modification of the original grain size distribution (\citealt{1996ApJ...469..740J}; \citealt{2011A&A...527A.123G}). However, the lower cutoff of the resulting grain size distribution is arbitrarily chosen to be $5$\AA~in the previous studies. 

On the other hand, thermal sputtering is believed a dominant process to destroy smallest grains such as PAHs and nanoparticles in hot shocked regions. Let $Y_{\rm sp}$ be the sputtering yield by a hydrogen atom bombarding on the dust grain at velocity $v_{\rm drift}$. The sputtering rate is given by
\bea
\frac{4\pi \rho a^{2}da}{dt} = n_{\H}v_{\rm drift}\pi a^{2}Y_{\rm sp}m_{\H},
\ena
which yields
\bea
\frac{da}{dt} &=& \frac{n_{\H}v_{\rm drift}m_{\H}Y_{\rm sp}}{4\rho}\nonumber\\
&\simeq&\left(\frac{0.1\mum}{\rm yr}\right)\left(\frac{v}{20\km\s^{-1}}\right)\hat{\rho}^{-1}n_{\H}Y_{\rm sp}.
\ena

For the C-shock where the neutral drift velocity $v_{\rm drift}< 50\km\s^{-1}$ (see Figure \ref{fig:Cshockn6-velo}), the sputtering yield is rather low, $Y_{\rm sp}< 10^{-3}$ \citep{1995Ap&SS.233..111D}. Therefore, the destruction time for grains of radius $a$ is given by
\bea
\tau_{\rm sp}&=&\frac{a}{da/dt}\simeq1.2\times 10^{3}\hat{\rho}a_{-7}n_{4}^{-1} \left(\frac{20\km\s^{-1}}{v_{\rm drift}}\right)\nonumber\\
&&\times\left(\frac{10^{-4}}{Y_{\rm sp}}\right) \rm yr,
\ena
where $n_{4}=n_{\H}/(10^{4}\cm^{-3})$. Comparing $\tau_{sp}$ with $\tau_{\rm disr}$ (Eq. \ref{eq:tdisr}), we can see that the disruption time is much shorter than the sputtering time for $a$ below several nanometers. 

In this paper, we identified a new mechanism of destruction for nanoparticles in the shocks, namely rotational disruption by stochastic mechanical torques. We showed that nanoparticles can be spun-up to extremely fast rotation due to high gas density, high temperature, and supersonic drift of neutrals relative to charged nanoparticles. As a result, nanoparticles are disrupted at some location in the shock where the tensile stress induced by centrifugal force exceeds the ultimate tensile strength. Physically, the high gas density and temperature due to shock compression allow gas collisional excitation to dominate over the electric dipole damping, making nanoparticles to rotate at thermal angular velocities. Supersonic neutral drift then further drive nanoparticles to suprathermal rotation. The rotational disruption is quite efficient for weak nanoparticles. For strong nanoparticles of tensile strength $S_{\rm max}=10^{10}\erg\cm^{-3}$, small nanoparticles of $a<1$ nm (e.g., PAHs) are removed from the C-shock at $v_{s}=30\km\s^{-1}$. We find that nanoparticles of ideal material of $S_{\max}=10^{11}\erg\cm^{-3}$ cannot be destroyed in C-shocks with $v_{s}\sim 30\km\s^{-1}$. Therefore, nanodiamonds produced in the previous stage of high velocity shock by grain-grain collision \citep{1987ApJ...319L.109T} would survive the shock and will be released in the ISM.

Table \ref{tab:destr} summarizes the destruction mechanisms in the shocks and their characteristic timescales. The grain-grain collision timescale is estimated as the mean time between two collisions, $\tau_{gg}=1/(\pi a^{2} n_{gr}v_{\rm drift})=4\rho aM_{g/d}/(3n_{\H}m_{\H}v_{\rm drift})$, assuming the single size $a$ distribution with the gas-to-dust mass ratio $M_{g/d}= 100$. Moreover, thermal sublimation is obviously ineffective in dense regions with low radiation intensity considered in this paper. The rotational disruption appears to be the fastest mechanism to destroy nanoparticles in C-shocks. As a result, this mechanism would play an important role in constraining the lower cutoff of grain size distribution in dense C-shock molecular clouds. 

It is noted that in the present study, we assumed spherical shapes for nanoparticles and only considered the rotational excitation by stochastic mechanical torques. Realistic nanoparticles are expected to be irregular, such that the regular mechanical torques are shown to be stronger (\citealt{2007ApJ...669L..77L}; \citealt{2018ApJ...852..129H}). As a result, the efficiency of mechanical disruption is perhaps more efficient than our present results.

\begin{table}
\begin{center}
\caption{Grain destruction in C-shocks}\label{tab:destr}
\begin{tabular}{l l} \hline\hline\\
{\it Mechanism} & {Timescales (yr)}\cr
\hline\\
Rotational disruption & $0.5 a_{-7}^{4}n_{4}^{-1}v_{\rm drift,1}^{-3}S_{\rm max,10}$ $^{a,b}$\cr
Thermal sputtering & $3.1\times 10^{3}\hat{\rho}a_{-7}n_{4}^{-1}T_{3}^{-1/2}({10^{-4}}/{Y_{\rm sp}})$ $^c$\cr
Non-thermal sputtering & $2.4\times 10^{3}\hat{\rho}a_{-7}n_{4}^{-1}v_{\rm drift,1}^{-1}({10^{-4}}/{Y_{\rm sp}})$\cr
Grain-grain collision & $7.6\times 10^{3}\hat{\rho}a_{-5}n_{4}^{-1}v_{\rm drift,1}^{-1}$\cr
\cr
\hline
\cr
\multicolumn{2}{l}{$^{a}$~$v_{\rm drift,1}=v_{\rm drift}/(10\rm km\s^{-1})$}\cr 
\multicolumn{2}{l}{$^{b}$~$S_{\rm max,10}=S_{\max}/(10^{10} \erg \cm^{-3})$}\cr
\multicolumn{2}{l}{$^c$~$T_{3}=T_{\gas}/(10^{3}\K)$} \cr
\hline\hline

\end{tabular}
\end{center}
\end{table}

\subsection{Effect of material tensile strength on rotational disruption}
The tensile strength of dust grains and nanoparticles is very uncertain. Ideal material such as graphene has highest tensile strength of $S_{\rm max}\sim 1.3 \times 10^{12}\erg\cm^{-3}$. If ideal PAHs can be considered as a sheet of graphene originating from collisions of graphite grains. Thus, one expect ideal PAHs to have a high tensile strength. Nevertheless, in astrophysical conditions, bombardment of energetic ions from cosmic rays can eject carbon atoms (see \citealt{2011A&A...526A..52M}), resulting in some defects. As a result, the tensile strength of PAHs can be reduced considerably, perhaps to $S_{\rm max}\sim 10^{9}\erg\cm^{-3}$.

Nanosilicate, nanoiron, and nanodiamond particles can have high tensile strength. For example, the tensile strength of graphite is $S_{\rm max}\sim 2\times 10^{10}\erg\cm^{-3}$ (\citealt{1972JMatS...7..239M}). { However, in shocked regions, nanoparticles can be heated to high temperatures of $T_{d}\sim 100\K$ due to collisional heating (see Eq. 25 in \citealt{Le:2019wo})}. As a result, its tensile strength is expected to be reduced (\citealt{Idrissi:2016fr}). We show that nanoparticles with $S_{\rm max}\lesssim 10^{9}\erg\cm^{-3}$ and size $a\lesssim2$ nm would be disrupted in C-shocks, while stronger nanoparticles can survive the shock passage.

\subsection{Implications for mid-IR emission from shocked regions}
Most observations of supernova remnants (SNRs) do not show mid-IR PAH emission features ($\lambda=3.3, 6.2, 7.7, 8.6, 11.3\mum$) \citep{2009ApJ...693..713S}, except observations for N132D by \cite{2006ApJ...653..267T}. Since PAHs are expected to be abundant in shocks due to shattering of carbonaceous grains, the lack of PAH emission implies that PAH destruction is efficient in the shocked regions (see \citealt{Kaneda:2011jd} for a review). On the other hand, observations show ubiquitous mid-IR emission features (e.g., $11\mum$ and $21\mum$) which are produced by hot very small grains ($a\sim 5-50$ nm) stochastically heated by UV photons \citep{2011ApJ...742....7A}.

We note that \cite{2010A&A...510A..36M} studied destruction of PAH molecules in interstellar shocks with velocity $v\sim 50-200\km\s^{-1}$. The authors found that PAHs are destroyed by sputtering due to proton and electron bombardment for $v>100\km\s^{-1}$, but they can survive passing the shock at lower velocities.

{In light of this study, small PAHs (i.e., $a<1$ nm) can be disrupted efficiently even at low shock velocities of $v< 50\km\s^{-1}$, but nanoparticles of $a\gtrsim 1$ nm can still survive the shock passage. Therefore, the proposed rotational disruption mechanism can successfully explain the lack of mid-IR PAH features and ubiquitious features at $9$ and $21\mum$ from very small grains\citep{Rho:2018ee}.}

\subsection{Tracing nanoparticles in shocked regions with spinning dust}
PAHs and nanoparticles are expected to be abundant in shocked regions due to grain-grain collisions (\citealt{1996ApJ...469..740J}; \citealt{2011A&A...527A.123G}). Nevertheless, to date, there is no observational method available to test this top-down formation mechanism of nanoparticles. By modeling microwave emission from spinning dust, we show that spinning dust emission is very strong in shocked regions due to rotational excitation by supersonic neutral drift. Even with a moderate abundance of Si or C in nanoparticles of $Y_{C}=Y_{Si}=5\%$, the spectral flux of spinning dust is still dominant over thermal dust at frequencies below $100$ GHz. Spinning dust emissivity is several orders of magnitude higher than thermal dust when rotational disruption is disregarded.\footnote{Note that the spinning dust emissivity for C-shocks is two orders of magnitude higher than in PDRs (\citealt{Hoang:2010jy}) if rotational disruption is not considered. In the absence of rotational disruption, the grain size distribution skewed to smaller sizes as in \cite{2011A&A...527A.123G} would produce even much stronger emissivity, which would be observed.}

Therefore, future radio observations with ALMA and ngVLA would be used to probe nanoparticles and test the different dust destruction mechanisms in C-shocks.  

\subsection{Constraining the shock velocity in dense regions with spinning dust}
C-shocks are ubiquitous in the ISM, especially in the dense, magnetized molecular clouds due to the effect of jets from young stars. The popular technique to probe slow shocks is through molecular emission lines (CO, SiO, and H$_{2}$). Here we suggest a new technique to trace shocks using continuum microwave emission from spinning dust. This technique is based on spinning dust mechanism which is most suitable in shocked regions due to both highly rotational excitation by hot gas, supersonic neutral drift, and a high abundance of nanoparticles presumably formed by grain-grain collisions.

\begin{figure}
\includegraphics[width=0.45\textwidth]{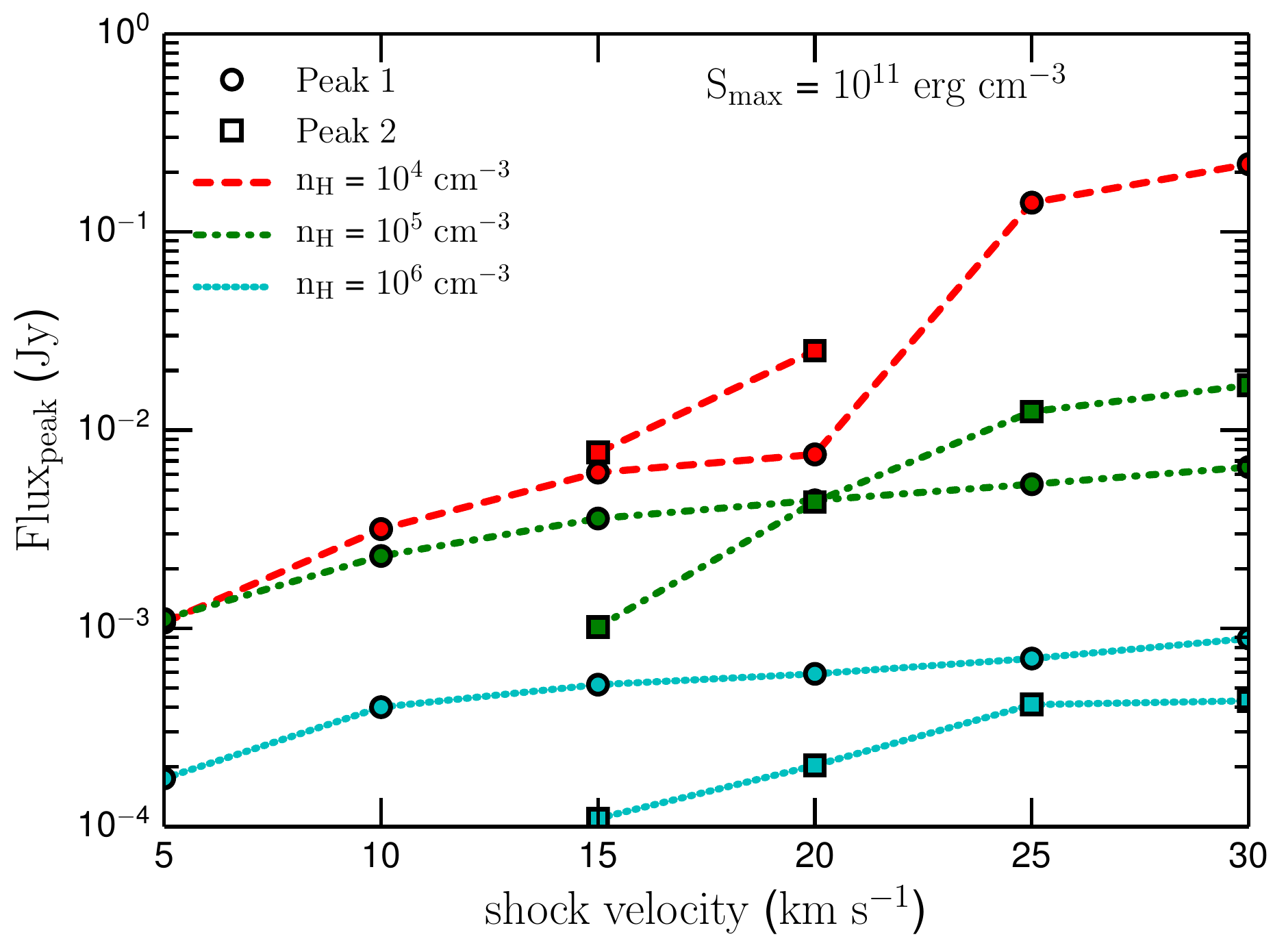}
\includegraphics[width=0.45\textwidth]{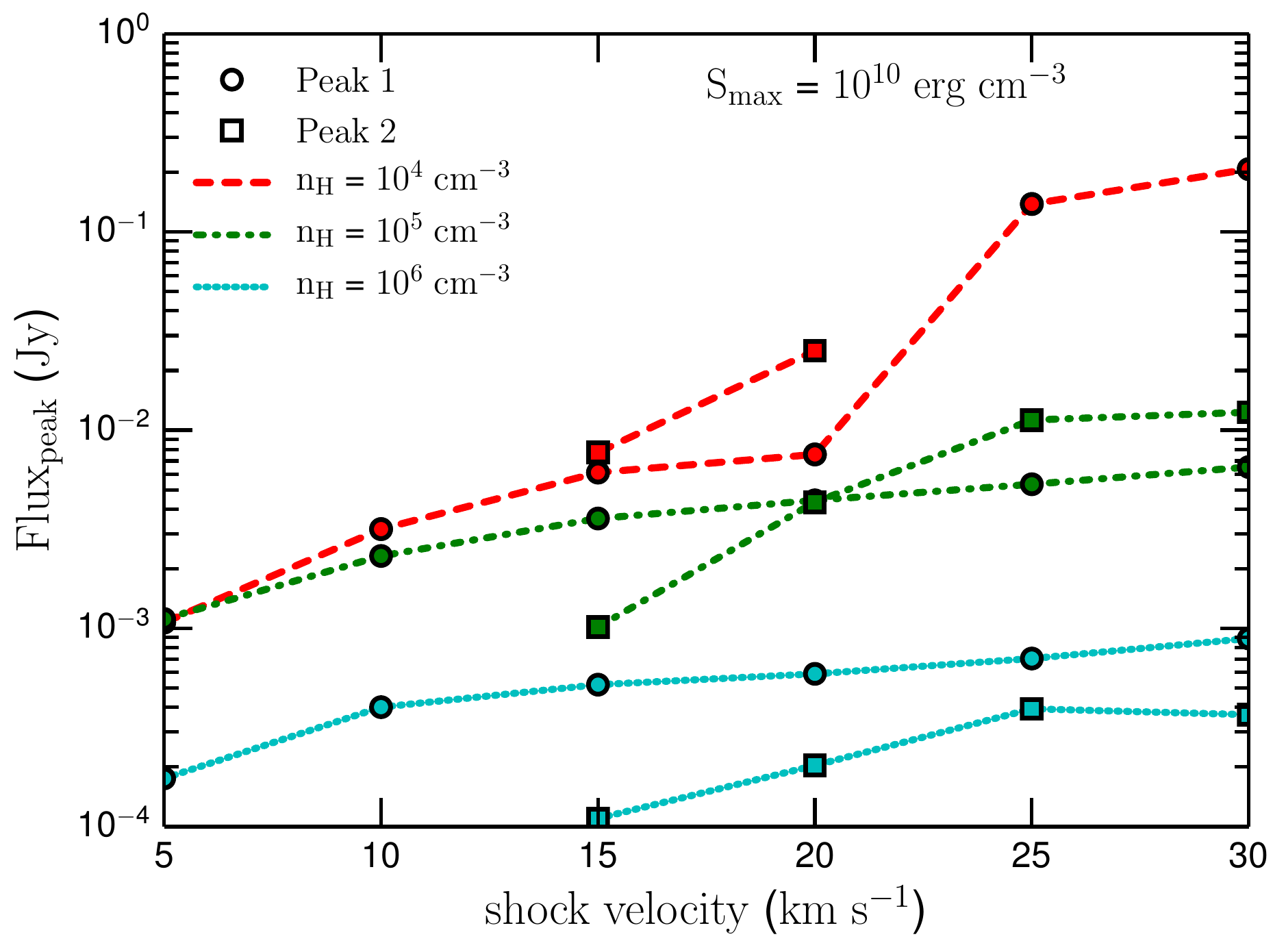}
\includegraphics[width=0.45\textwidth]{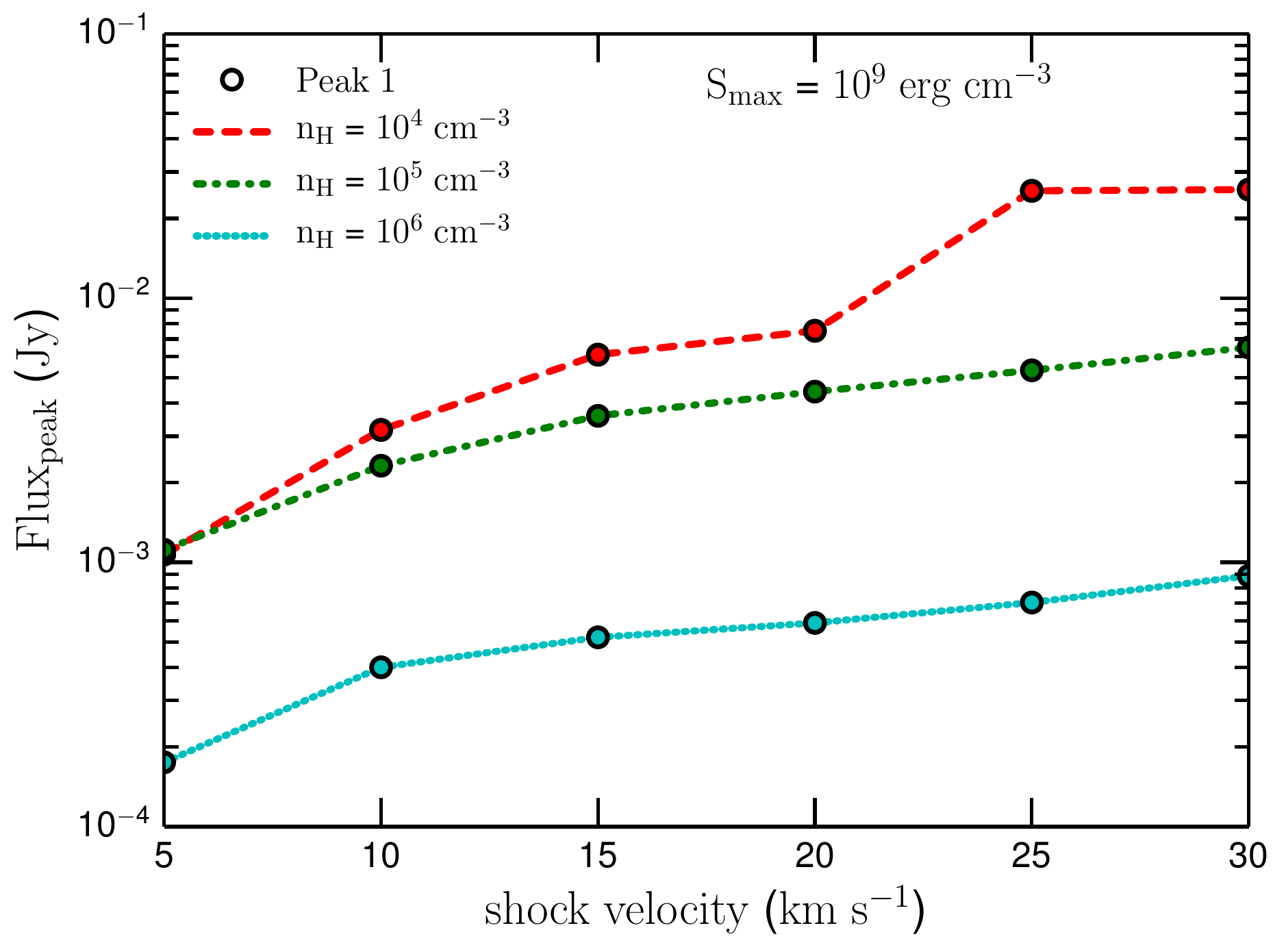}
\caption{Emission flux of the first maximum (peak 1) and second maximum (peak 2) of emission spectrum vs. shock velocity for three different shock models, $n_{\H}=10^{4}\cm^{-3}$, $n_{\H}=10^{5}\cm^{-3}$, and $n_{\H}=10^{6}\cm^{-3}$. Upper, middle, and lower panels show the results for three values of the tensile strength $S_{\rm max}$, respectively. The peak flux increases rapidly with increasing $v_{\rm s}$ until rotational disruption occurs.}
\label{fig:fluxmax_omega}
\end{figure}

Figure \ref{fig:fluxmax_omega} illustrates the variation of maximum emission flux (peak flux) of spinning dust emission spectrum with the shock velocity, which are obtained from Figures \ref{fig:flux_S10} and \ref{fig:flux_S9}. The peak flux increases rapidly with shock velocity due to both increased neutral drift velocity and higher gas temperature. The peak flux increases by a factor of 2 when the shock velocity increases from $v_{s}=15\km\s^{-1}$ to $30\km\s^{-1}$. In particular for the model A ($n_{\H}=10^{4}\cm^{-3}$, the peak flux increases significantly for $v_{s}\ge 20\km\s^{-1}$. For weak nanoparticles of $S_{\max}=10^{9}\erg\cm^{-3}$, there exists only one prominent peak (peak 1) because the second peak is destroyed due to the rotational disruption effect.

\begin{figure}
\includegraphics[width=0.45\textwidth]{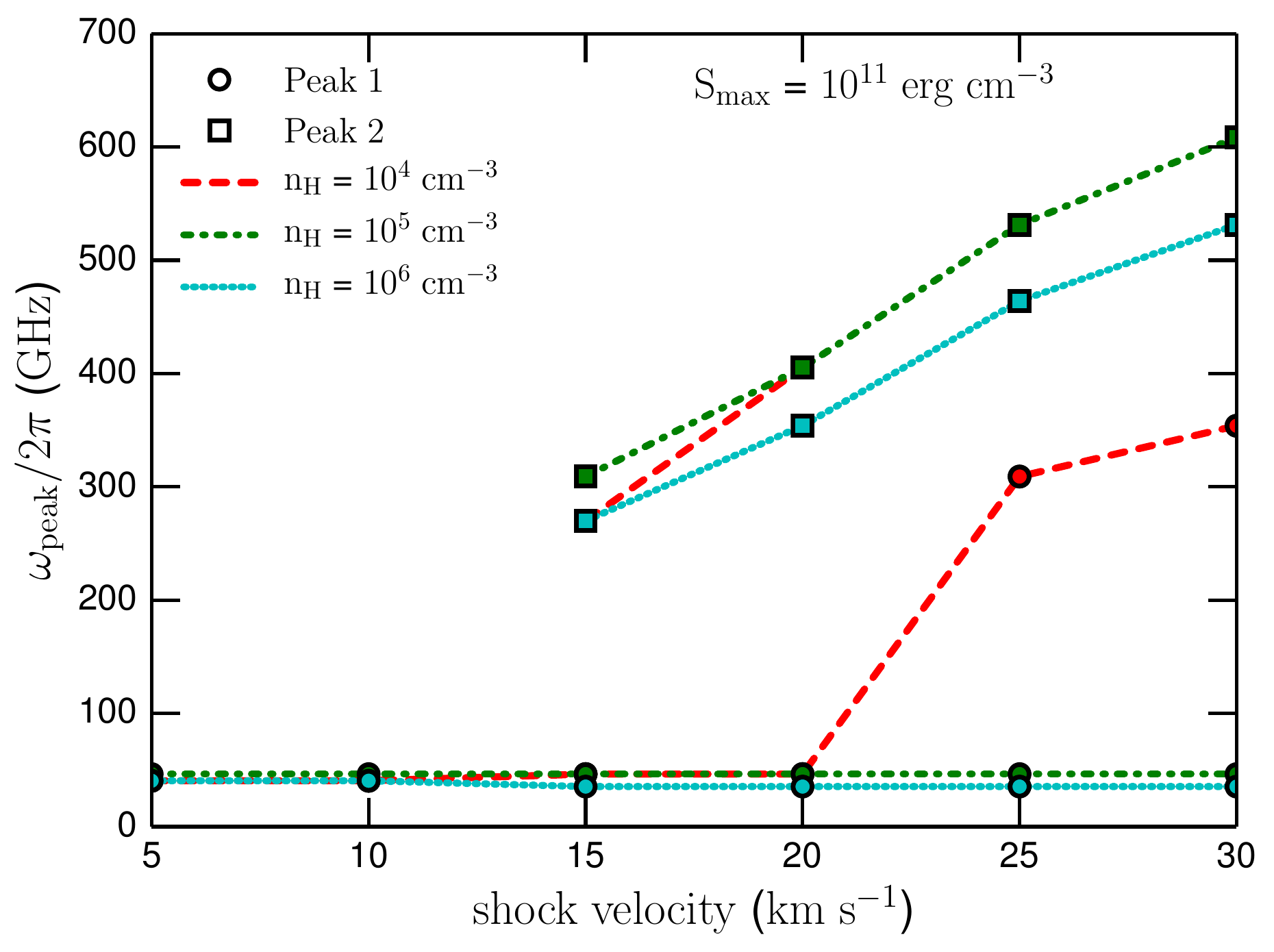}
\includegraphics[width=0.45\textwidth]{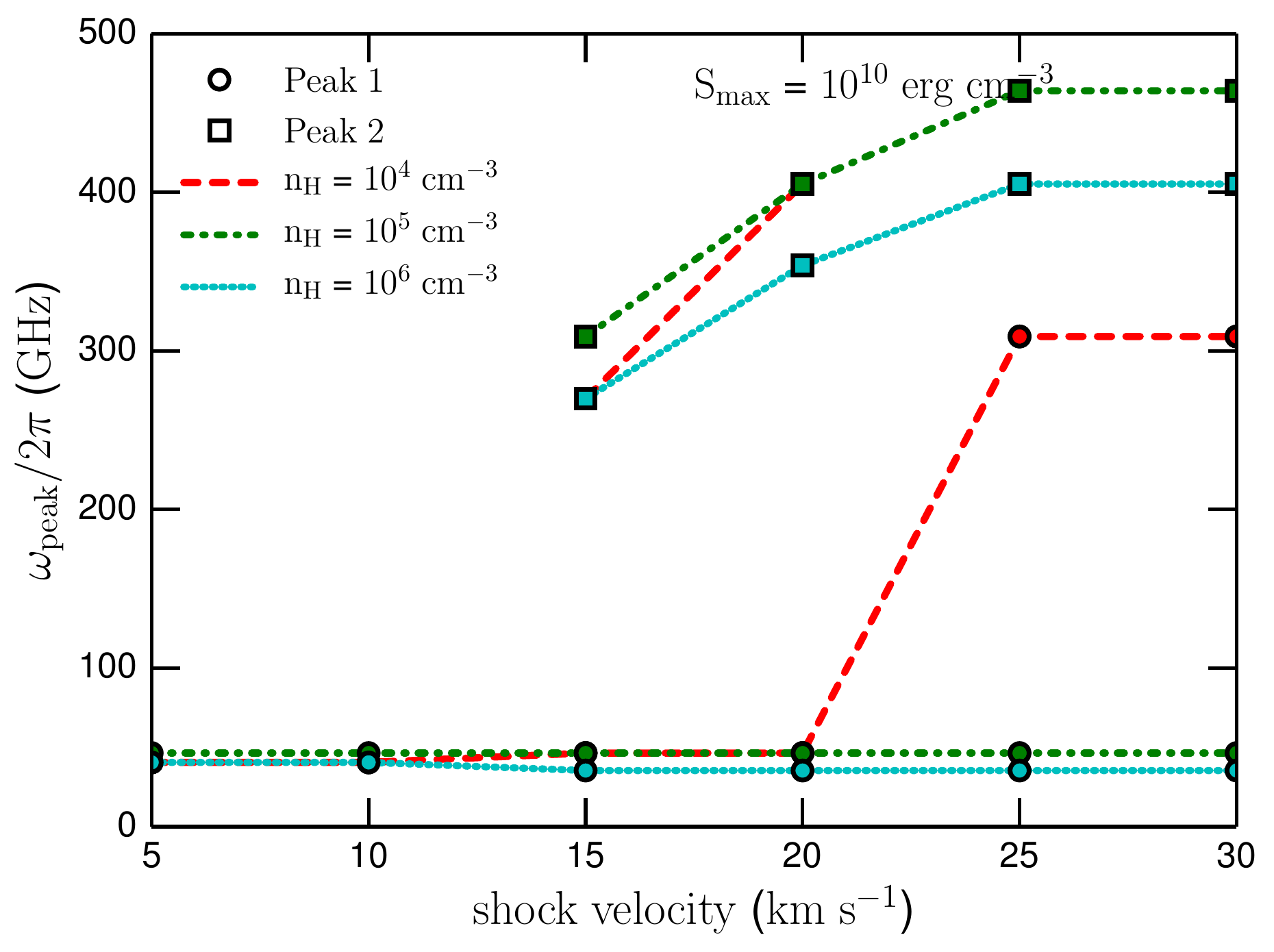}
\includegraphics[width=0.45\textwidth]{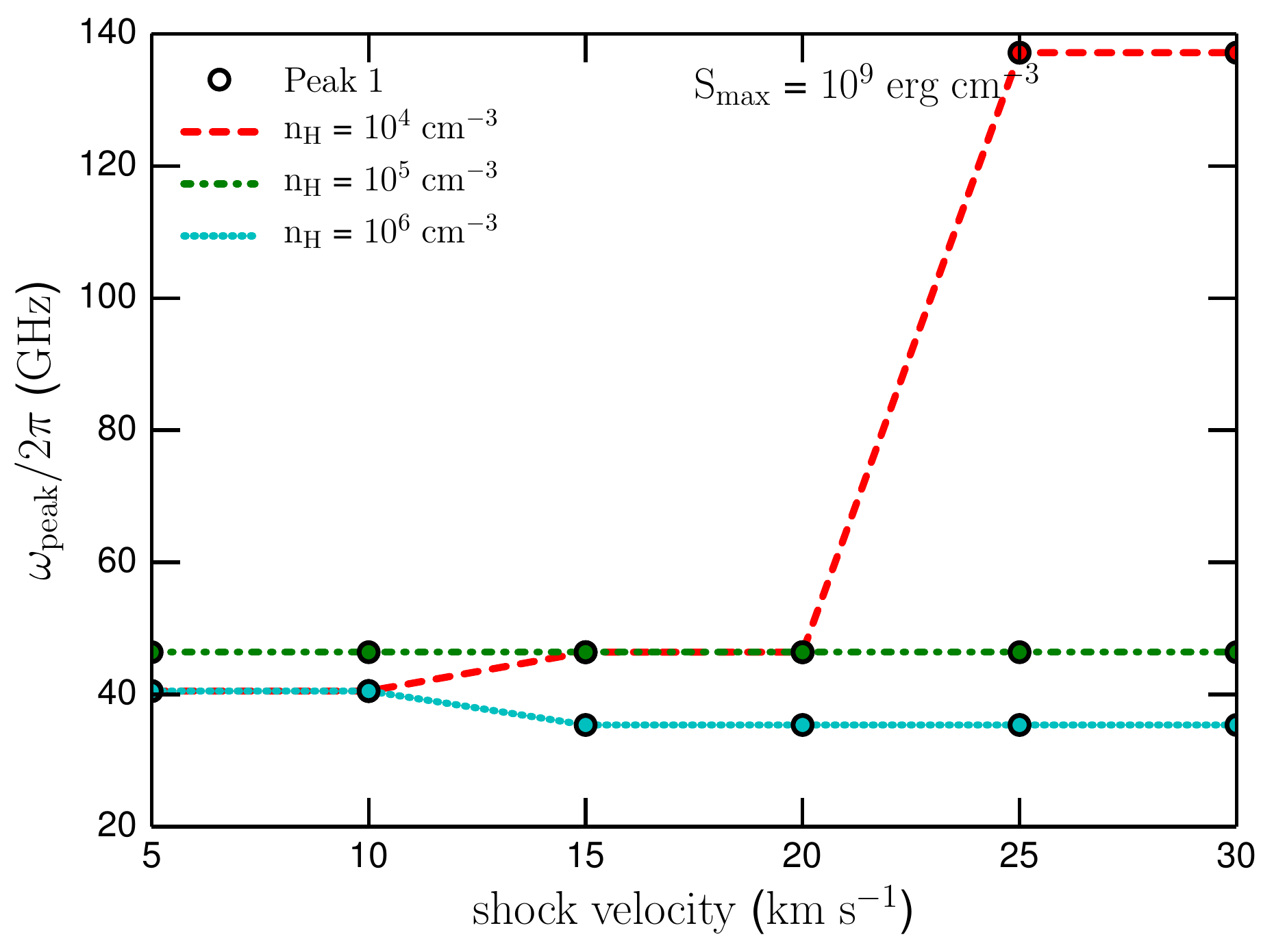}
\caption{Same as Figure \ref{fig:fluxmax_omega}, but for the peak emission frequency. The first peak frequency slightly changes with $v_{s}$ except for the model of $n_{\H}=10^{4}\cm^{-3}$, but the second peak frequency increases rapidly with $v_{s}$.}
\label{fig:omegamax_vs}
\end{figure}

Figure \ref{fig:omegamax_vs} shows the peak frequency as a function of shock velocity. 
The first peak frequency increases with increasing $v_{s}$ for model A ($n_{\H}=10^{4}\cm^{-3}$) only, but it slightly changes with $v_{s}$ for the other models (B and C) due to the enhanced minimum size $a_{\rm min}$ due to rotational disruption. On the other hand, the second peak frequency increases rapidly with $v_{s}$ for strong nanoparticles which are not disrupted by rotational disruption (upper and middle panels).

\subsection{Grain alignment and tracing magnetic fields in shocks}
Magnetic fields play a crucial role in regulating shock structures in dense clouds (\citealt{1980ApJ...241.1021D}). Dust polarization resulting from grain alignment is a powerful technique to trace magnetic fields. To date, a detailed study of grain alignment in shocked regions is not yet available. We found that grains can be spun-up to suprathermal rotation in shocks by stochastic torques. Grains of irregular shape experience also regular mechanical torques and can be aligned with the magnetic field (\citealt{2007ApJ...669L..77L}; \citealt{2016MNRAS.457.1958D}; \citealt{2018ApJ...852..129H}). If grains have iron inclusions which can greatly enhance its magnetic relaxation, grains can be aligned perfectly with the magnetic field (\citealt{2016ApJ...831..159H}). Detailed studies of grain alignment and resulting dust polarization in shocks will be presented elsewhere in a near future.

\section{Summary}\label{sec:sum}
We study rotational dynamics of nanoparticles in C-shocks of dense molecular clouds and discover a new mechanism of dust destruction by stochastic mechanical torques. Our principal results are summarized as follows:
\begin{itemize}

\item[1] For the first time, we study rotation dynamics of nanoparticles in C-shocks passing dense clouds, taking into account supersonic drift of neutral relative to charged nanoparticles. We find that charged nanoparticles can be rapidly spun-up to suprathermal rotation by stochastic bombardment of gas atoms. 

\item[2] We find that suprathermally rotating nanoparticles can be rapidly disrupted into tiny fragments because the centrifugal stress induced by grain rotation can exceed the maximum tensile strength of grain material. Nanoparticles made of made of weak materials ($S_{\max}\lesssim 10^{9}\erg\cm^{-3}$) are easier to be disrupted than nanoparticles of strong materials.

\item[3] We compare the characteristic timescale of rotational disruption with other destruction mechanisms and find that rotational disruption is the most efficient mechanism in C-shocks with sufficiently large velocities. Thus, the minimum size of nanoparticles is constrained by the rotational disruption instead of thermal sputtering. This rotational disruption mechanism can play an important role in dust destruction in dense and hot regions compressed by supernova shocks and reverse shocks.

\item[4] We perform modeling of microwave emission from rapidly spinning nanoparticles in C-shocks where the minimum size of nanoparticles is determined by rotational disruption. Even in the presence of rotational disruption, spinning dust emission is still dominant over thermal dust emission at frequencies $\nu<100$ GHz. We find that the peak frequency and emissivity increases with increasing shock velocity. 

\item[5] We suggest spinning dust emission as a new probe of nanoparticles and shock velocities in dense molecular clouds where nanoparticles are expected to be abundant due to grain-grain collisions and are rotating suprathermally due to excitation by supersonic neutral drift. 

\end{itemize}
\acknowledgments
{We thank the anonymous referees for comments that helped improve the presentation of our manuscript. We thank V. Guillet, A. Gusdorf, P. Lesaffre for various discussions and comments.} This work was supported by the Basic Science Research Program through the National Research Foundation of Korea (NRF), funded by the Ministry of Education (2017R1D1A1B03035359).



\bibliography{ms.bbl}

\end{document}